\documentclass[zpreprint,zbstepj]{./zeus_paper}
\usepackage[english]{babel}
\newcommand{\ZcoosysB}{%
The ZEUS coordinate system is a right-handed Cartesian system, with the $Z$
axis pointing in the proton beam direction, referred to as the ``forward
direction'', and the $X$ axis pointing left towards the centre of HERA.
The coordinate origin is at the nominal interaction point.\xspace}
\newcommand{\Zpsrap}{%
The pseudorapidity is defined as $\eta=-\ln\left(\tan\frac{\theta}{2}\right)$,
where the polar angle, $\theta$, is measured with respect to the proton beam
direction.\xspace}

\chardef\usc=95
\chardef\til=126
\catcode`\@=11 
\DeclareRobustCommand\xdotspace{\futurelet\@let@token\@xdotspace}
\def\@xdotspace{%
  \ifx\@let@token.\else
  \ifx\@let@token\bgroup.\else
  \ifx\@let@token\egroup.\else
  \ifx\@let@token\/.\else
  \ifx\@let@token\ .\else
  \ifx\@let@token~.\else
  \ifx\@let@token!.\else
  \ifx\@let@token,.\else
  \ifx\@let@token:.\else
  \ifx\@let@token;.\else
  \ifx\@let@token?.\else
  \ifx\@let@token/.\else
  \ifx\@let@token'.\else
  \ifx\@let@token).\else
  \ifx\@let@token-.\else
  \ifx\@let@token\@xobeysp.\else
  \ifx\@let@token\space.\else
  \ifx\@let@token\@sptoken.\else
   .\space
   \fi\fi\fi\fi\fi\fi\fi\fi\fi\fi\fi\fi\fi\fi\fi\fi\fi\fi}
\catcode`\@=12 

\newcommand{\stru}[2]{%
   \relax\ifmmode\hbox{\vrule height#1 depth#2 width0pt}%
   \else\vrule height#1 depth#2 width0pt\fi}

\newcommand{\Ronum}[1]{\uppercase\expandafter{\romannumeral#1}}
\newcommand{\ronum}[1]{\expandafter{\romannumeral#1}}
\DeclareRobustCommand{\LaTeXZ}{%
  \LaTeX\kern-.05em4\kern-.1em
  {\raisebox{-0.2ex}{$\scriptstyle\text{ZEUS}$}}\xspace}

\DeclareMathAlphabet{\mathbf}{OT1}{cmr}{bx}{sl}
\newcommand{\eVdist}{\kern-0.06667em}

\newcommand{\Gev}{{\text{Ge}\eVdist\text{V\/}}}

\newcommand{\Tesla}{\,\text{T}}

\newcommand{\slashfrac}[2]{%
  \raisebox{0.5ex}{\ensuremath #1}\kern-0.12em/\kern-0.08em
  \raisebox{-.8ex}{\ensuremath #2}}

\newcommand{\sqr}[3]{%
    {\vcenter{\hrule height.#3ex\hbox{\vrule width.#2ex height#1ex
     \kern#1ex\vrule width.#3ex}\hrule height.#2ex}}}

\newcommand{\widebar}[1]{%
   \mkern1.5mu\overline{\mkern-1.5mu#1\mkern-1.mu}\mkern1.mu}
\catcode`\@=11 
\newcommand{\parenbar}{\mathpalette\p@renb@r}
\def\p@renb@r#1#2{\vbox{%
  \ifx#1\scriptscriptstyle \dimen@.7em\dimen@ii.2em\else
  \ifx#1\scriptstyle \dimen@.8em\dimen@ii.25em\else
  \dimen@1em\dimen@ii.4em\fi\fi \offinterlineskip
  \ialign{\hfill##\hfill\cr
    \vbox{\hrule width\dimen@ii}\cr
    \noalign{\vskip-.3ex}%
    \hbox to\dimen@{$\mathchar300\hfil\mathchar301$}\cr
    \noalign{\vskip-.3ex}%
    $#1#2$\cr}}}
\catcode`\@=12 

\newcommand{\qbar}{\widebar{q}}

\newcommand{\als}{\alpha_s}

\newcommand{\IP}{{\rm I$\kern-0.01667em$P}\xspace}

\mathchardef\qsm=63
\mathchardef\pls=43
\mathchardef\mns=512
\mathchardef\plm=518
\mathchardef\eql=61
\mathchardef\smallleft=300
\mathchardef\smallright=301
\mathchardef\les=316
\mathchardef\gre=318
\mathchardef\leq=532
\mathchardef\grq=533

\catcode`\@=11 
\newcounter{pict@width}
\newcounter{pict@height}
\newlength{\pict@scale}
\newcommand{\psfigadd}[4]{%
\setcounter{pict@width}{1*\ratio{#2+\pict@scale/2}{\pict@scale}}
\setcounter{pict@height}{1*\ratio{#3+\pict@scale/2}{\pict@scale}}
\setlength{\unitlength}{\pict@scale}
\hbox to #2{\hspace{-\fill}\begin{picture}(\thepict@width,\thepict@height)
\put(0,0){\psfig{figure=#1,width=#2,height=#3,clip=}}
\SetScale{0.283466457}
\SetWidth{1.763889}
{#4}
\end{picture}}
}
\newcounter{pict@widthfst}
\newcounter{pict@widthscd}
\newcounter{pict@widthtot}
\newcommand{\psfigaddtwo}[7]{%
\setcounter{pict@widthfst}{1*\ratio{#2+\pict@scale/2}{\pict@scale}}
\setcounter{pict@widthscd}{1*\ratio{#2+#4+\pict@scale/2}{\pict@scale}}
\setcounter{pict@widthtot}{1*\ratio{#2+#4+#6+\pict@scale/2}{\pict@scale}}
\setcounter{pict@height}{1*\ratio{#3+\pict@scale/2}{\pict@scale}}
\setlength{\unitlength}{\pict@scale}
\hbox{\hspace{-\fill}\begin{picture}(\thepict@widthtot,\thepict@height)
\put(0,0){\psfig{figure=#1,width=#2,height=#3,clip=}}
\put(\thepict@widthscd,0){\psfig{figure=#5,width=#6,height=#3,clip=}}
\SetScale{0.283466457}
\SetWidth{1.763889}
{#7}
\end{picture}}
}
\newcommand{\psfigror}[4]{%
\setcounter{pict@width}{1*\ratio{#2+\pict@scale/2}{\pict@scale}}
\setcounter{pict@height}{1*\ratio{#3+\pict@scale/2}{\pict@scale}}
\setlength{\unitlength}{\pict@scale}
\hbox{\begin{picture}(\thepict@width,\thepict@height)
\put(0,\thepict@height){\psfig{figure=#1,width=#3,height=#2,clip=,angle=270}}
\SetScale{0.283466457}
\SetWidth{1.763889}
{#4}
\end{picture}}
}
\newcommand{\psfigrol}[4]{%
\setcounter{pict@width}{1*\ratio{#2+\pict@scale/2}{\pict@scale}}
\setcounter{pict@height}{1*\ratio{#3+\pict@scale/2}{\pict@scale}}
\setlength{\unitlength}{\pict@scale}
\hbox{\begin{picture}(\thepict@width,\thepict@height)
\put(0,0){\psfig{figure=#1,width=#3,height=#2,clip=,angle=90}}
\SetScale{0.283466457}
\SetWidth{1.763889}
{#4}
\end{picture}}
}
\catcode`\@=12 
\newlength\listtextwidth

\catcode`\@=11 
\newlength{\@tabfninsert}
\newlength{\@tabfnwidth}
\newcommand{\tabfootnote}[2]{%
  \setlength{\@tabfninsert}{0.8em}
  \setlength{\@tabfnwidth}{\textwidth}
  \addtolength{\@tabfnwidth}{-\@tabfninsert}
  \addtolength{\@tabfnwidth}{-0.4em}
  \noindent\makebox[\@tabfninsert][r]{\footnotesize$^{#1}$\hfil}\hfill%
  \parbox[t]{\@tabfnwidth}{\footnotesize #2\hfill}}
\catcode`\@=12 

\newcommand{\ptmiss}{{\mbox{$\not\hspace{-.3ex}{p}_T$}}}
\newcommand{\nsub}{$\langle n_{\rm sbj}\rangle\;$}

\def\citeCTD{{\cite{%
nim:a279:290,*npps:b32:181,*nim:a338:254%
}}\xspace}
\def\citeCAL{{\cite{%
nim:a309:77,*nim:a309:101,*nim:a321:356,*nim:a336:23%
}}\xspace}

\begin{document}
\prepnum{{DESY--03--055}}

\title{
Jet production  in charged current deep inelastic $e^+p$ scattering at HERA
}

\author{ZEUS Collaboration}

\date{June, 2003}
                     
\abstract{
The production rates and substructure of jets have been studied in charged
current deep inelastic $e^+p$ scattering for $Q^2>200$~GeV$^2$  with
the ZEUS detector at HERA using an integrated luminosity of 110.5~pb$^{-1}$. 
Inclusive jet cross
sections are presented for jets with transverse energies $E_T^{\rm jet} > 14$~GeV
and pseudorapidities in the range $-1 < \eta^{\rm jet}< 2$. Dijet 
cross sections are presented for events with a jet having $E_T^{\rm jet} > 14$~GeV and  a second jet having $E_T^{\rm jet} > 5$~GeV. Measurements of the mean subjet multiplicity, $\langle n_{\rm sbj}\rangle$, of the inclusive jet sample are presented. 
Predictions based on parton-shower Monte Carlo models and next-to-leading-order QCD calculations are compared to the measurements. 
The value of $\als(M_Z)$, determined from \nsub at $y_{\rm cut}=10^{-2}$ for jets with $25<E_T^{\rm jet}<119$ GeV, is $\als(M_Z) \;=\; 0.1202 \;\pm 0.0052 \;({\rm stat.}) \;^{+0.0060}_{-0.0019} \;\;({\rm syst.})\; ^{+0.0065}_{-0.0053}\;({\rm th.})$. The mean subjet multiplicity as a function of $Q^2$ is found to be consistent with that measured in NC DIS.
} 

\makezeustitle

\pagenumbering{Roman}                                                                              
                                                   %
\begin{center}                                                                                     
{                      \Large  The ZEUS Collaboration              }                               
\end{center}                                                                                       
  S.~Chekanov,                                                                                     
  M.~Derrick,                                                                                      
  D.~Krakauer,                                                                                     
  J.H.~Loizides$^{   1}$,                                                                          
  S.~Magill,                                                                                       
  B.~Musgrave,                                                                                     
  J.~Repond,                                                                                       
  R.~Yoshida\\                                                                                     
 {\it Argonne National Laboratory, Argonne, Illinois 60439-4815}~$^{n}$                            
\par \filbreak                                                                                     
  M.C.K.~Mattingly \\                                                                              
 {\it Andrews University, Berrien Springs, Michigan 49104-0380}                                    
\par \filbreak                                                                                     
  P.~Antonioli,                                                                                    
  G.~Bari,                                                                                         
  M.~Basile,                                                                                       
  L.~Bellagamba,                                                                                   
  D.~Boscherini,                                                                                   
  A.~Bruni,                                                                                        
  G.~Bruni,                                                                                        
  G.~Cara~Romeo,                                                                                   
  L.~Cifarelli,                                                                                    
  F.~Cindolo,                                                                                      
  A.~Contin,                                                                                       
  M.~Corradi,                                                                                      
  S.~De~Pasquale,                                                                                  
  P.~Giusti,                                                                                       
  G.~Iacobucci,                                                                                    
  A.~Margotti,                                                                                     
  R.~Nania,                                                                                        
  F.~Palmonari,                                                                                    
  A.~Pesci,                                                                                        
  G.~Sartorelli,                                                                                   
  A.~Zichichi  \\                                                                                  
  {\it University and INFN Bologna, Bologna, Italy}~$^{e}$
% ...
\newpage                                                                                           
$^{\    1}$ also affiliated with University College London \\                                      
$^{\    2}$ on leave of absence at University of                                                   
Erlangen-N\"urnberg, Germany\\                                                                     
% ...
                                                           %
                                                           %
\newpage   
                                                           %
                                                           %
\begin{tabular}[h]{rp{14cm}}                                                                       
$^{a}$ &  supported by the Natural Sciences and Engineering Research                               
          Council of Canada (NSERC) \\                                                             
$^{b}$ &  supported by the German Federal Ministry for Education and                               
          Research (BMBF), under contract numbers HZ1GUA 2, HZ1GUB 0, HZ1PDA 5, HZ1VFA 5\\         
$^{c}$ &  supported by the MINERVA Gesellschaft f\"ur Forschung GmbH, the                          
          Israel Science Foundation, the U.S.-Israel Binational Science                            
          Foundation and the Benozyio Center                                                       
          for High Energy Physics\\                                                                
$^{d}$ &  supported by the German-Israeli Foundation and the Israel Science                        
          Foundation\\                                                                             
$^{e}$ &  supported by the Italian National Institute for Nuclear Physics (INFN) \\                
$^{f}$ &  supported by the Japanese Ministry of Education, Culture,                                
          Sports, Science and Technology (MEXT) and its grants for                                 
          Scientific Research\\                                                                    
$^{g}$ &  supported by the Korean Ministry of Education and Korea Science                          
          and Engineering Foundation\\                                                             
$^{h}$ &  supported by the Netherlands Foundation for Research on Matter (FOM)\\                   
$^{i}$ &  supported by the Polish State Committee for Scientific Research,                         
          grant no. 620/E-77/SPUB-M/DESY/P-03/DZ 247/2000-2002\\                                   
$^{j}$ &  partially supported by the German Federal Ministry for Education                         
          and Research (BMBF)\\                                                                    
$^{k}$ &  supported by the Fund for Fundamental Research of Russian Ministry                       
          for Science and Edu\-cation and by the German Federal Ministry for                       
          Education and Research (BMBF)\\                                                          
$^{l}$ &  supported by the Spanish Ministry of Education and Science                               
          through funds provided by CICYT\\                                                        
$^{m}$ &  supported by the Particle Physics and Astronomy Research Council, UK\\                   
$^{n}$ &  supported by the US Department of Energy\\                                               
$^{o}$ &  supported by the US National Science Foundation\\                                        
$^{p}$ &  supported by the Polish State Committee for Scientific Research,                         
          grant no. 112/E-356/SPUB-M/DESY/P-03/DZ 301/2000-2002, 2 P03B 13922\\                    
$^{q}$ &  supported by the Polish State Committee for Scientific Research,                         
          grant no. 115/E-343/SPUB-M/DESY/P-03/DZ 121/2001-2002, 2 P03B 07022\\                    
\end{tabular}                                                                                      
                                                           %
                                                           %

\pagenumbering{arabic} 
\pagestyle{plain}

\section{Introduction}
\label{sec-int}

Measurements of the charged current (CC) deep inelastic scattering (DIS) cross section at HERA~\cite{pl:b324:241,*prl:75:1006,*zfp:c67:565,*pl:b379:319,zfp:c72:47,epj:c12:411,*pl:b539:197,epj:c13:609,*epj:c19:269} at high virtuality, $Q^2$, of the exchanged boson have demonstrated the presence of a space-like propagator with a finite mass, consistent with that of the $W$ boson.
Jet production in CC DIS provides a testing ground for  QCD as well as the electroweak sector of the Standard Model. 
Up to leading order in the strong coupling constant, $\als$, jet production in CC DIS proceeds via the QCD-Compton ($Wq \to q^{\prime}g$) and W-gluon-fusion ($Wg \to q\qbar^{\prime}$) processes in addition to the pure electroweak process ($Wq \to q^{\prime}$).

At HERA, multijet structure has been observed in CC DIS~\cite{zfp:c72:47,epj:c19:429} at large $Q^2$ and jet substructure has been studied using the differential and integrated jet shapes~\cite{epj:c8:367}. Another useful representation of the internal jet structure  is the 
subjet multiplicity~\cite{np:b383:419,*pl:b378:279,np:b421:545}.
 The lowest-order non-trivial contribution to the subjet multiplicity is of 
order $\als$, so that measurements of the subjet multiplicity provide a direct test of QCD.

This paper reports a detailed study of the hadronic final state in CC $e^+p$ DIS. Differential cross sections are presented for both inclusive jet and dijet production. The jets were identified in the laboratory frame using the longitudinally invariant $k_T$ cluster algorithm~\cite{np:b406:187}.
After describing experimental conditions and the theoretical calculations, in Section~\ref{incjet} the inclusive jet cross sections are presented as a function of the virtuality of the exchanged boson, the jet pseudorapidity, $\eta^{\rm jet}$, and the  jet transverse energy, $E_T^{\rm jet}$.
In Section~\ref{dijet}, the dependence of the dijet cross sections on $Q^2$ and the invariant mass, $m_{12}$, of the two highest-$E_T$ jets are given. In Section~\ref{subjet}, the mean subjet multiplicity, $\langle n_{\rm sbj}\rangle,$ as a function of the resolution scale, $y_{\rm cut}$ and $E_T^{\rm jet}$ using the inclusive jet sample is presented.  Parton-shower Monte Carlo (MC) calculations and 
next-to-leading-order (NLO) QCD predictions~\cite{ijmp:a4:1781} are compared to the measurements. 
In Section~\ref{alphas}, the value of $\als(M_Z)$ determined using the measurements of \nsub as a function of $E_T^{\rm jet}$ is given.
In Section~\ref{subjetccnc}, the measurements of \nsub as a function of $E_T^{\rm jet}$ and $Q^2$ are compared to the results obtained by ZEUS in neutral current (NC) DIS~\cite{misc:zeus:eps01:641}.

\section{Experimental conditions}
\label{sec-exp}

The data sample used in this analysis was collected with the ZEUS detector at HERA and corresponds to an integrated luminosity of 110.5 pb$^{-1}$. During the 1995-1997 (1999-2000) running period,  HERA operated with protons of energy $E_p=820$~GeV (920~GeV) and positrons of energy $E_e=27.5$~GeV, yielding a centre-of-mass energy of 300~GeV (318~GeV).  
A detailed description of the ZEUS detector can be found 
elsewhere~\cite{zeus:1993:bluebook}. A brief outline of the 
components that are most relevant for this analysis is given
below. Charged particles are tracked in the central tracking detector (CTD)~\citeCTD,
which operates in a magnetic field of $1.43\Tesla$ provided by a thin 
superconducting solenoid. The CTD consists of 72~cylindrical drift chamber 
layers, organized in nine superlayers covering the polar-angle\footnote{{\ZcoosysB\Zpsrap}} region \mbox{$15^\circ<\theta<164^\circ$}. The transverse-momentum resolution for
full-length tracks is $\sigma(p_T)/p_T=0.0058p_T\oplus0.0065\oplus0.0014/p_T$,
with $p_T$ in $\Gev$. The high-resolution uranium-scintillator calorimeter (CAL)~\citeCAL consists 
of three parts: the forward (FCAL), the barrel (BCAL) and the rear (RCAL)
calorimeters. Each part is subdivided transversely into towers and
longitudinally into one electromagnetic section (EMC) and either one (in RCAL)
or two (in BCAL and FCAL) hadronic sections (HAC). The smallest subdivision of
the calorimeter is called a cell.  The CAL energy resolutions, as measured under
test-beam conditions, are $\sigma(E)/E=0.18/\sqrt{E}$ for electrons and
$\sigma(E)/E=0.35/\sqrt{E}$ for hadrons, with $E$ in $\Gev$. Jet energies were corrected for the energy lost in inactive material, typically about one radiation length, in front of the CAL. The effects of the uranium noise were minimised by discarding cells in the electromagnetic or hadronic sections if they had energy deposits of less than 60 MeV or 110 MeV, respectively. A three-level trigger~\cite{zeus:1993:bluebook,trigger2} was used to select events online. 

The luminosity was measured using the Bethe-Heitler reaction $e^+p \to e^+p\gamma$. The resulting small-angle energetic photons were measured by the luminosity monitor~\cite{Desy-92-066,*zfp:c63:391,*acpp:b32:2025}, a lead-scintillator calorimeter placed in the HERA tunnel at $Z = -107$ m.

\section{Data selection and jet search}
\label{sec-data}

The selection of charged current events for the present study is very
similar to those described in detail in previous ZEUS publications~\cite{epj:c12:411,*pl:b539:197}. The efficiency of the selection cuts is typically above 90$\%$ and the remaining backgrounds are negligible. 

The principal signature of a CC DIS event at HERA is the presence of a large missing transverse momentum, $\ptmiss$, arising from the energetic final-state neutrino which escapes detection. The quantity $\ptmiss$ was calculated from 

\centerline{$ \ptmiss^2 = p_X^2+p_Y^2= \left ( \displaystyle\sum_{i} E_i \sin\theta_i  \cos\phi_i \right )^2 +  \left( \displaystyle\sum_{i} E_i \sin\theta_i  \sin\phi_i\right )^2 $ , }

where the sums run over all CAL cells, i, $E_i$ is the energy deposit and $\theta_i$, $\phi_i$ are the polar and azimuthal angles of the cell  as viewed from the interaction vertex. 
The total transverse energy, $E_T$, is given by $E_T = \sum E_i \sin\theta_i$.
   
The inelasticity, $y$, 
was reconstructed using the Jacquet-Blondel method~\cite{proc:epfacility:1979:391} and corrected for detector effects as described previously~\cite{zfp:c72:47}. 
The detector simulation was used to derive corrected values {{\mbox{$\not\hspace{-.3ex}{p}_{T,{\rm cor}}$}} and $y_{\rm cor}$. The corrected value of $Q^2$, $Q^2_{\rm cor}$, was calculated in terms of {{\mbox{$\not\hspace{-.3ex}{p}_{T,{\rm cor}}$}} and $y_{\rm cor}$ using the relation $Q^2_{\rm cor}=$ {{\mbox{$\not\hspace{-.3ex}{p}_{T,{\rm cor}}^{\;2}$}} $/(1-y_{\rm cor})$.

The following requirements were imposed on the data sample:
\begin{itemize}
\item $\ptmiss > 11$ GeV and $Q^2_{\rm cor} > 200$ GeV$^2$, to ensure high trigger efficiency;
\item $y_{\rm cor} < 0.9$, to avoid the degradation of the resolution in $Q^2$ near $y\sim 1$;
\item $\ptmiss / E_T > 0.5$, to reject photoproduction and beam-gas background. For the dijet sample, this cut was reduced to $\ptmiss / E_T > 0.3$ with the further requirement that the difference between the azimuthal angle of the missing transverse momentum and that of the closest jet was greater than 1 rad. This cut removed poorly reconstructed back-to-back dijet photoproduction  events;
\item a vertex position reconstructed with the CTD in the range $-50 < Z < 50$ cm, consistent with an $ep$ interaction;
\item the difference, $\Delta \phi$,  between the azimuthal angle of the net transverse momentum as measured by the tracks associated with the vertex and that measured from the CAL 
be  less than $1$ rad. This requirement removed random coincidences of cosmic rays with $ep$ interactions;
\item $p_T^{\rm track}/\ptmiss >0.1$, where $p_T^{\rm track}$ is the net transverse momentum of the tracks associated with the vertex. This condition was not applied if $\ptmiss > 25$ GeV. This cut rejected events with additional energy deposits in the CAL not related to $ep$ interactions (mainly cosmic rays) and beam-related background in which $\ptmiss$ has a small polar angle;

\item the event was removed from the sample if there was an isolated positron candidate with energy above 10 GeV, to reject NC DIS events;
\item a pattern-recognition algorithm based on the topology of the calorimeter energy distribution and the signals detected in the muon chambers was applied to reject cosmic rays and beam-halo muons. 

\end{itemize}

The longitudinally invariant $k_T$ cluster algorithm~\cite{np:b406:187} was used in the inclusive mode~\cite{pr:d48:3160} to reconstruct jets in the hadronic final state both in data and in MC simulated events (see Section~\ref{MC}). In data and MC, the algorithm was applied to the energy deposits in the CAL cells and in the MC it was also applied to the final-state hadrons. The jet search was performed in the $\eta-\phi$ plane of the laboratory, starting with the CAL cells or hadrons as initial objects. In the following discussion, $E_{T,i}$ denotes the transverse
energy, $\eta_i$ the pseudorapidity and $\phi_i$ the azimuthal angle of
object $i$ in the laboratory frame. For each pair of objects, the quantity
\begin{equation}
 d_{ij} = [(\eta_i - \eta_j)^2 + (\phi_i - \phi_j)^2 ] \cdot {\min}(E_{T,i},E_{T,j})^2 \nonumber
\end{equation}
was calculated. For each individual object, the quantity $d_i = (E_{T,i})^2$
was also calculated. If, of all the values $\{d_{ij},d_i \}$, $d_{kl}$ was
the smallest, then objects $k$ and $l$ were combined into a single new
object. If, however, $d_k$ was the smallest, then object $k$ was
considered a jet and was excluded from further clustering. The procedure was
repeated until all objects were assigned to jets. The jet variables were
defined according to the Snowmass convention~\cite{proc:snowmass:1990:134}:
\begin{equation*}
E_T^{\rm jet}  = \displaystyle\sum_i E_{T,i}\; , \; 
\eta^{\rm jet} = \displaystyle\sum_i \frac{E_{T,i}\eta_i}{E_T^{\rm jet}}\; , \;
 \phi^{\rm jet}=\displaystyle\sum_i \frac{E_{T,i}\phi_i}{E_T^{\rm jet}},
\end{equation*}
where the sums run over all objects associated with the given jet.
This prescription was also used to determine these variables for the subjets.
 For jets constructed from CAL cells, jet energies were corrected for all energy-loss effects, principally in inactive material of typically about one radiation length, in front of the CAL (see Section~\ref{MC}).

For the inclusive jet sample, all jets with $E_T^{\rm jet} > 14$ GeV and $-1 <  \eta^{\rm jet} < 2$ were retained. For the dijet sample, 
 at least one additional jet with $E_T^{\rm jet}>5$ GeV and $-1 <  \eta^{\rm jet} < 2$ was required. The upper rapidity requirement
is made so that the jet is within the CTD acceptance for efficient background
rejection.  There are very few events with jets with sufficient $E_T^{\rm jet}$ below the lower rapidity requirement.

With the above criteria, 1865 events with at least one jet and 282 dijet events were identified.

\subsection{Definition of subjet multiplicity}
\label{subsec-defsub}
Subjets were resolved within a jet by considering all objects associated with the jet and by repeating the application of the $k_T$ cluster algorithm described above, until for every pair of objects $i$ and $j$, the quantity $d_{ij}$ was greater than $d_{\rm cut} = y_{\rm cut} \cdot \left ( E_T^{\rm jet}\right )^2$. All remaining objects were called subjets. The jet structure depends upon the value chosen for the resolution parameter $y_{\rm cut}$. For each sample studied, the mean subjet multiplicity, $\langle n_{\rm sbj}\rangle$,  is defined as the average number of subjets contained  in a jet at a given value of $y_{\rm cut}$:
\begin{equation}
\label{meansub}
\langle n_{\rm sbj}(y_{\rm cut})\rangle = \frac{1}{N_{\rm jets}}\displaystyle\sum_{i=1}^{N_{\rm jets}} n^{i}_{\rm sbj}(y_{\rm cut}), \nonumber
\end{equation} 
where $n^{i}_{\rm sbj}$ is the number of subjets in jet $i$ and $N_{\rm jets}$ is the total number of jets in the sample. The mean subjet multiplicity of the inclusive jet sample was measured for $y_{\rm cut}$ values in the range $5 \cdot 10^{-4}$ to $0.1$. The $y_{\rm cut}$ range was chosen to be small enough to have mean subjet multiplicities larger than unity and large enough to avoid the degradation in resolution caused by the finite size of the CAL cells.   
\section{Monte Carlo simulation}
\label{MC}
Samples of events were generated to determine the response of the detector to jets of hadrons and to evaluate the correction factors necessary to obtain the hadron-level  jet cross sections and subjet multiplicities. 
The CC DIS events were generated using the LEPTO 6.5 program~\cite{cpc:101:108} interfaced to HERACLES 4.6.1~\cite{cpc:69:155,*spi:www:heracles} via DJANGOH 1.1~\cite{cpc:81:381,*spi:www:djangoh11}. 
The HERACLES program includes first-order electroweak radiative corrections. The CTEQ4D~\cite{pr:d55:1280} NLO proton parton distribution functions (PDF) were used.
The QCD radiation was modelled with the colour-dipole model~\cite{pl:b165:147,*pl:b175:453,*np:b306:746,*zp:c43:625} by using the ARIADNE 4.08 program~\cite{cpc:71:15,*zp:c65:285} including the boson-gluon-fusion process.  As an alternative, samples of events were generated using the LEPTO model which is based on first-order QCD matrix elements and parton showers. For the generation of the LEPTO samples, the option for soft-colour interactions was switched off since its inclusion results in an increase both in particle multiplicity and energy per unit of rapidity that disagrees with the measurements in NC DIS at HERA~\cite{epj:c11:251}. In both cases, fragmentation into hadrons was performed using the Lund string model~\cite{prep:97:31} as implemented in JETSET 7.4~\cite{cpc:39:347,*cpc:43:367}. 
 To calculate the acceptances and to estimate hadronisation effects, the generated events were passed through the GEANT 3.13-based~\cite{tech:cern-dd-ee-84-1} simulation of the ZEUS detector and trigger.
They were reconstructed and analysed by the same program chain as used for data. 
For both the ARIADNE and LEPTO event samples, a good description of the measured distributions for the kinematic and jet variables was obtained~\cite{moni_thesis}. 

To correct the data to hadron level, multiplicative correction factors, defined as the ratio of the measured quantities for jets of hadrons over the same quantity for jets at detector level, were estimated by using the ARIADNE and LEPTO models. Parton-level predictions were also obtained by applying the jet algorithm to the MC-generated partons. These predictions were used to correct the NLO QCD calculations to hadron level (Section~\ref{nlo}). 

HERACLES 4.6.2~\cite{cpc:69:155,*spi:www:heracles} was used to correct the measured cross sections to the electroweak Born level evaluated using the electromagnetic coupling constant $\alpha = 1/137.03599$, the Fermi coupling constant $G_F = 1.16639\cdot 10^{-5} GeV^{-2}$ and the mass of the Z boson $M_Z=91.1882$ GeV~\cite{epj:c15:1} to determine the electroweak parameters. 

\section{NLO QCD calculations}
\label{nlo}
The NLO QCD calculations were obtained from the program MEPJET~\cite{pl:b380:205}, which employs the phase-space slicing method~\cite{pr:d46:1980}. This is the only available program providing NLO calculations for jet production in charged current deep inelastic scattering. 
 The calculations were performed in the $\overline{MS}$ renormalisation  and factorisation  schemes. The number of flavours was set to five and the renormalisation $(\mu_R)$ and factorisation $(\mu_F)$ scales were chosen to be $\mu_R=\mu_F=Q$. The calculations were performed using the CTEQ4M~\cite{pr:d55:1280} parametrisations of the proton PDFs, which are based on the $\overline{MS}$ scheme.  The jet algorithm described in Section~\ref{sec-data} was also applied to the partons in the events generated by MEPJET in order to compute the jet cross section and the predictions for the subjet multiplicities.  The cross sections were evaluated using the same values for $\alpha$, $G_F$ and $M_Z$ as in the electroweak Born level of the measured cross sections (Section~\ref{MC}). In addition, the mass of the W boson was fixed to $80.4603$ GeV.

Since the measurements correspond to jets of hadrons whereas the NLO QCD calculations correspond to jet of partons, the predictions were corrected to the hadron level using the MC simulations. The multiplicative correction factor ($C_{\rm had}$) is defined as the ratio of either the cross sections or the mean subjet multiplicities for jets of hadrons to the same quantity for jets of partons, estimated using the MC programs described in Section~\ref{MC}. The ratios obtained with the ARIADNE and LEPTO models were in good agreement and the mean was taken as the value of $C_{\rm had}$.
The value of $C_{\rm had}$ is $\sim 1.03 \;(\sim 1.10)$ for the inclusive jet (dijet) cross sections. 
For the mean subjet multiplicity, $C_{\rm had}$ is $2.13$ at $y_{\rm cut}=5\cdot 10^{-4}$ and $14 < E_T^{\rm jet} < 17$ GeV and approaches unity as $y_{\rm cut}$ and $E_T^{\rm jet}$ increase.

The theoretical predictions were redetermined after changing the
parameters as described below. In each case the difference between the
redetermination and the nominal prediction was taken to be the
uncertainty in the calculation associated with the parameter under
consideration.

\begin{itemize}

\item Proton PDFs: 
the CTEQ5M~\cite{epj:c12:375} and MRST~\cite{epj:c4:463,*epj:c14:133} sets, rather than 
CTEQ4M~\cite{pr:d55:1280} , were used. Also, a set of the MRST PDFs with a larger $d/u$ quark 
ratio at large Bjorken $x$ was used. The uncertainty in the cross sections was less than $\sim 4\%$ for the inclusive jet cross section, except for high $E_T^{\rm jet}$, where it reaches $\sim 20\%$. It was less than $\sim 10\%$ for the dijet cross sections. The uncertainty was negligible for the subjet multiplicities;

\item $\als(M_Z)$: the  $\als(M_Z)$ values of $0.113$ and $0.119$,  corresponding to the proton PDFs 
CTEQ4A2 and CTEQ4A4, were used. The uncertainty in the cross sections
was typically $\sim 2\%$; for the mean subjet multiplicity the
uncertainty was $\sim 1\%$ ;

\item $\mu_R$: in order to estimate the effects of the terms beyond NLO, the scale $\mu_R$ was varied between $Q/2$ and $2Q$, while keeping $\mu_F$ fixed at $Q$. The uncertainty of the cross sections was less than $5\%$. The uncertainty on the  mean subjet multiplicity was $\sim 3\%$ for $y_{\rm cut}=10^{-2}$;

\item $C_{\rm had}$: the hadronisation correction, $C_{\rm had}$, was varied by half of the
 difference between those evaluated using ARIADNE and LEPTO.  The
 uncertainty typically amounted to less than 1\%(3\%) for inclusive jet
 (dijet) cross sections. For the subjet multiplicities, the
 uncertainty was less than $3\%$ for $y_{\rm cut}=10^{-2}$.

\item $s_{\rm min}$: the cut-off parameter $s_{\rm min}$ in the phase-space slicing was changed from the default value of 0.1 GeV$^2$ to 0.01 GeV$^2$. This uncertainty was less than $1\%$ in all the calculations and was neglected in the estimation of the total theoretical uncertainty.

\end{itemize}

The total theoretical uncertainty was obtained by adding in quadrature
the individual uncertainties listed above and is shown as the hatched
band in the figures.

\section{Experimental systematic uncertainties}
\label{syst}
A study of the sources contributing to the systematic uncertainties of the measurements was carried out. The following sources were considered:

\begin{itemize}
\item the uncertainty on the absolute energy scale of the jets was taken to be $\pm 1\%$ for $E_T^{\rm jet}>10$ GeV and $\pm 3\%$ for lower $E_T^{\rm jet}$ values~\cite{pl:b531:9,*epj:c23:615,*conf_wing}. The resulting uncertainty was less than 5$\%$ (12 $\%$) for the inclusive jet (dijet) cross sections and less than 2$\%$ for the mean subjet multiplicity; 

\item the uncertainty in the reconstruction of the kinematic variables due to that in the absolute energy scale of the CAL was estimated by varying the energy variables measured with the CAL by  $\pm 3\%$.  The uncertainty was less than 5$\%$ for all distributions;

\item  the differences in the results obtained by using ARIADNE or LEPTO to correct the data for detector effects were taken as systematic uncertainties; they were typically smaller than 5$\%$ for the cross sections and smaller than 2$\%$ for the mean subjet multiplicities;

\item the selection cut of $\ptmiss > 11$ GeV was changed to $10$ GeV and $12$ GeV. This gave a variation of the cross sections (subjet multiplicities) of less than  5$\%$ $(2\%)$. The uncertainty evaluated from the variation of other selection cuts was typically less than $2 \%$.
\end{itemize}

For the jet cross sections, the systematic uncertainties not associated with the absolute energy scale of the jets and the CAL are not point-to-point correlated and were added in quadrature to the statistical errors. They are shown as the bars in the figures. The uncertainty due to the absolute energy scale is point-to-point correlated and is shown separately as a shaded band in each figure. For the subjet multiplicities all the systematic uncertainties are point-to-point correlated and were added in quadrature to the statistical errors. They are shown as the bars in the figures.

In addition, there is an overall normalisation uncertainty of $2.0\%$ from the luminosity determination, which is not included in the results presented in the figures and the tables of the cross sections.


\section{Results}

\subsection{Data-combination method}
\label{subsec-datacomb}

Due to the different centre-of-mass energy of the two data sets used in the analysis, the measured jet cross sections based on each set are presented separately in Tables~\ref{table-q2incjet} to~\ref{table-m12dijet}. 

The measured jet cross sections, $\sigma_{\sqrt{s}}$, were combined using the following formula:

\begin{equation*}\label{equa_comb}
\sigma^{\rm comb}_{318}= \frac{\sigma_{300} \cdot {\cal{L}}_{300} + \sigma_{318} \cdot {\cal{L}}_{318}} {{\cal{L}}_{300} \cdot ({\sigma_{300}^{th}}/{\sigma_{318}^{th}}) + {\cal{L}}_{318}},
\end{equation*}
where ${\cal{L}}_{\sqrt{s}}$ is the luminosity  and  $\sigma_{\sqrt{s}}^{th}$ is the predicted cross section. The ratio ${\sigma_{300}^{th}}/{\sigma_{318}^{th}}$ was obtained using the program MEPJET. The ratio
obtained by ARIADNE is within $1 \%$ of that obtained by MEPJET and
was the same, within statistical errors, as that measured in the data. All the systematic errors have been assumed to be correlated between the measurements.

No dependence of the subjet multiplicities with the centre-of-mass energy was seen either in the data or in the theoretical predictions; thus the subjet multiplicities were calculated using the inclusive jet sample of both data sets. The measured subjet multiplicities are presented in Tables~\ref{table-sbjycut1} to~\ref{table-sbq2}.

\subsection{Inclusive jet differential cross sections}
\label{incjet}

The differential inclusive jet cross sections were measured in the kinematic region defined by $Q^2>200$ GeV$^2$ and $y<0.9$. These cross sections include every jet of hadrons in the event with $E_T^{\rm jet}>14$ GeV and $-1<\eta^{\rm jet}<2$. The differential inclusive jet cross sections as a function of $Q^2$, $\eta^{\rm jet}$ and $E_T^{\rm jet}$ are shown in Figs.~\ref{fig-incq2} to~\ref{fig-incet}. Both the ARIADNE MC model and the NLO QCD calculation MEPJET give a good description of the measured inclusive jet cross sections.

\subsection{Dijet differential cross sections}
\label{dijet}

The differential dijet cross sections were measured in the kinematic region defined by $Q^2>200$ GeV$^2$ and $y<0.9$. These cross sections refer to the two jets of hadrons with highest transverse energy in the event with $E_T^{{\rm jet},1}>14$ GeV, $E_T^{{\rm jet},2}>5$ GeV and $-1<\eta^{\rm jet}<2$. The differential dijet cross sections as a function of $Q^2$ and  the invariant mass of the two highest-$E_T$ jets, $m_{12}$, are presented in Figs.~\ref{fig-diq2} and \ref{fig-dim12}. The  NLO QCD calculation gives a good description of the measured dijet cross sections.

\subsection{Subjet multiplicities}
\label{subjet}
The mean subjet multiplicity, $\langle n_{\rm sbj}\rangle$, was determined using the inclusive sample of jets in the kinematic region defined by $Q^2>200$ GeV$^2$ and $y<0.9$. The \nsub values were obtained using every jet of hadrons in the event with $E_T^{\rm jet}>14$ GeV and $-1<\eta^{\rm jet}<2$. 
The results are shown in Fig.~\ref{fig-subyet} as a function of $y_{\rm cut}$ for different $E_T^{\rm jet}$ regions. In the region of small $y_{\rm cut}$, the ARIADNE MC model gives a better description of the multiplicity than the NLO QCD calculation MEPJET. At larger values of $y_{\rm cut}$ both the MC model and the NLO QCD calculation give a good description of the measurement. In the region of small $y_{\rm cut}$ values, fixed-order QCD calculations are  affected by large uncertainties and a resummation of terms enhanced by $ln(y_{\rm cut})$~\cite{np:b421:545,*jhep:b9909:009} would be required for a precise comparison with the data. At 
relatively large values of $y_{\rm cut}$, a NLO fixed-order calculation is expected~\cite{np:b421:545,*jhep:b9909:009} to be a good approximation to such a resummed calculation.

The measured \nsub at $y_{\rm cut}=10^{-2}$ as a function of $E_T^{\rm jet}$ is shown in Fig.~\ref{fig-alphassent}. 
The measured mean subjet multiplicity decreases as $E_T^{\rm jet}$ increases. The overall description of the data by the NLO QCD calculations is good.

\section{Measurement of $\als$}
\label{alphas}

The sensitivity of the subjet multiplicity to the value of $\als(M_Z)$ is illustrated in Fig.~\ref{fig-alphassent}, which compares the measured \nsub at $y_{\rm cut}=10^{-2}$ as a function of $E_T^{\rm jet}$ with NLO QCD calculations obtained with different values of $\als(M_Z)$.  Both the measurements and the NLO QCD predictions of the subjet multiplicities have smaller uncertainties
compared to those of the jet cross sections.
Therefore, the measured $\langle n_{\rm sbj}\rangle$, rather than the jet cross sections,
was used to determine $\als(M_Z)$ using the following procedure:
\begin{itemize}
\item the NLO QCD calculations of \nsub were performed for the five sets of the CTEQ4 ``A-series'' PDFs, which differ in the assumed value of $\als(M_Z)$. The value of $\als(M_Z)$ used in each calculation was that associated with the corresponding set of PDFs;
\item for each bin $i$ in $E_T^{\rm jet}$, the NLO QCD calculations, corrected for hadronisation effects, were used to parametrise the $\als(M_Z)$ dependence of \nsub according to

\begin{equation}
\label{fitalphas}
\left [  \langle n_{\rm sbj}\rangle \; (\als(M_Z)) \; \right ]_i \;= 1 + \;C_1^i\; \; \als(M_Z) \;+\;  C_2^i \;  \als(M_Z)^2.
\end{equation} 

The coefficients $C_1^i$ and $C_2^i$ were determined by performing a $\chi^2$-fit to the NLO QCD predictions. This simple parametrisation gives a good description of the $\als(M_Z)$ dependence of \nsub over the entire range spanned by the CTEQ4 ``A-series'';

\item this parametrisation was used to extract a value of $\als(M_Z)$ in each bin;
\item in addition, a combined value of $\als(M_Z)$ was determined by a $\chi^2$-fit of Eq.~(\ref{fitalphas}) to the measured \nsub values for all bins. 
\end{itemize}

This procedure correctly handles the complete $\als$ dependence of the calculations (the explicit dependence coming from the partonic cross sections as well as the implicit dependence coming from the PDFs) in the fit, while preserving the correlation between $\als$ and the PDFs.

The uncertainty in the extracted values of $\als(M_Z)$ due to the experimental systematic uncertainties was evaluated by repeating the above analysis for each systematic check. The largest contribution to the experimental uncertainty was that due to the simulation of the hadronic final state.

The theoretical uncertainties arising from terms beyond NLO and uncertainties in the hadronisation correction were evaluated as described in Section~\ref{nlo}. These resulted in uncertainties in $\als(M_Z)$ of $\Delta\als(M_Z)=^{+0.0064}_{-0.0051}$ and $\Delta\als(M_Z)=\pm 0.0014$, respectively. The total theoretical uncertainty was obtained by adding these in quadrature. Other uncertainties described in Section~\ref{nlo} were small and were neglected.  As a cross check, a linear parametrisation of the $\als(M_Z)$ dependence of \nsub was considered; the change in the extracted value of $\als(M_Z)$ was negligible.

The values of $\als(M_Z)$ obtained from the measurement of \nsub for various $E_T^{\rm jet}$ regions are in good agreement. The value of $\als(M_Z)$ obtained from the measurements of \nsub at $y_{\rm cut}=10^{-2}$ for $25<E_T^{\rm jet}<119$ GeV, a region in which the parton-to-hadron correction was less than $10\%$, is
\newline

\centerline{$\als(M_Z) \;=\; 0.1202 \;\pm 0.0052 \;({\rm stat.}) \;^{+0.0060}_{-0.0019} \;\;({\rm syst.})\; ^{+0.0065}_{-0.0053}\;({\rm th.})$.}

This result is consistent with other recent determinations using measurements in NC DIS of inclusive jet~\cite{oscar_inc,epj:c19:289} and exclusive dijet cross sections~\cite{pl:b507:70} as well as measurements of $\langle n_{\rm sbj}\rangle$~\cite{misc:zeus:eps01:641} and with the PDG value, $\als(M_Z)=0.1172\pm0.0020$~\cite{new_pdg}.

\section{Comparison of subjet multiplicities in CC and NC}

\label{subjetccnc}

The present measurements of subjet multiplicities in CC interactions are compared with the corresponding measurements in NC DIS~\cite{misc:zeus:eps01:641}. The NC data were reanalysed in the same kinematic region as that of the CC analysis. 

The measurements of \nsub at the value of $y_{\rm cut}=10^{-2}$ as a function of $E_T^{\rm jet}$ in CC and NC DIS are compared in Fig.~\ref{fig-subyetnc}a. The value of \nsub is slightly larger for jets in NC DIS than for CC DIS for a given jet transverse energy. 
The NLO QCD predictions behave in the same way as the data.

The subprocess population and the phase space available for QCD radiation depend on $Q^2$. The  measurements of \nsub at $y_{\rm cut}=10^{-2}$ as a function of $Q^2$ in CC and NC DIS are compared in Fig.~\ref{fig-subyetnc}b. The values of \nsub in CC and NC DIS are similar and are in agreement with the NLO predictions. The differences observed in the subjet multiplicity as a function of $E_T^{\rm jet}$ can be attributed to the different $Q^2$ distributions of the CC and NC processes.

\section{Summary}
Measurements of differential cross sections for inclusive jet and dijet production in charged current deep inelastic $e^+p$ scattering have been performed and are corrected to the electroweak Born level. The internal structure of the inclusive jet sample has been studied in terms of the mean subjet multiplicity. The results are given for jets of hadrons identified with the longitudinally invariant $k_T$ cluster algorithm in the laboratory frame in the kinematic region defined by $Q^2>200$ GeV$^2$ and $y<0.9$.  Inclusive jet cross
sections are presented for jets with transverse energies $E_T^{\rm jet} > 14$~GeV
and pseudorapidities in the range $-1 < \eta^{\rm jet}< 2$. Dijet 
cross sections are presented for events with a jet having $E_T^{\rm jet} > 14$~GeV and  a second jet having $E_T^{\rm jet} > 5$~GeV.

The predictions of the ARIADNE MC model and NLO QCD calculations obtained with the program MEPJET give a good description of the measurements of inclusive
and dijet cross sections.
 
The average number of subjets decreases as $E_T^{\rm jet}$ increases. 
The NLO QCD calculations agree well with the measured subjet multiplicities, $\langle n_{\rm sbj}\rangle$.  A fit of the measured \nsub as a function of $E_T^{\rm jet}$ at $y_{\rm cut}=10^{-2}$ provides a determination of the strong coupling constant $\als(M_Z)$. The value of $\als(M_Z)$ determined for the region $E_T^{\rm jet}>25$ GeV is
\newline

\centerline{$\als(M_Z) \;=\; 0.1202 \;\pm 0.0052 \;({\rm stat.}) \;^{+0.0060}_{-0.0019} \;\;({\rm syst.})\; ^{+0.0065}_{-0.0053}\;({\rm th.})$.}

This result is consistent with other recent determinations and with the PDG value.

The subjet multiplicities in CC and NC DIS are similar as a function of $Q^2$. 
The measured \nsub at a given $E_T^{\rm jet}$ is somewhat smaller in CC DIS than in NC DIS. This can be attributed to the different $Q^2$ distributions of the two processes.

\section{Acknowledgements}

We thank the DESY Directorate for their strong support and encouragement. The remarkable achievements of the HERA machine group were essential for the successful completion of this work and are greatly appreciated. We are grateful for the support of the DESY computing and network services. The design, construction and installation of the ZEUS detector have been made possible owing to the ingenuity and effort of many people who are not listed as authors. We would like to thank D.~Zeppenfeld for useful discussions and help in running his program for calculating QCD jet cross sections in charged current interactions.

\pagestyle{plain}
{\raggedright
\providecommand{\etal}{et al.\xspace}
\providecommand{\coll}{Collaboration}
\catcode`\@=11
\def\@bibitem#1{%
\ifmc@bstsupport
  \mc@iftail{#1}%
    {;\newline\ignorespaces}%
    {\ifmc@first\else.\fi\orig@bibitem{#1}}
  \mc@firstfalse
\else
  \mc@iftail{#1}%
    {\ignorespaces}%
    {\orig@bibitem{#1}}%
\fi}%
\catcode`\@=12

}
\vfill\eject

\begin{table}
\begin{center}
\begin{tabular}{|l|cccc||c|c|}
\hline
\multicolumn{7}{|c|}{1995-1997 $e^+p$ data sample ($\sqrt{s}=300$ GeV)}\\
\hline
\multicolumn{1}{|c|}{\raisebox{0.25cm}[1.cm]{\parbox{2cm}{\centerline{$Q^2$ range} \centerline{(GeV$^2$)}}}}
          & \raisebox{0.25cm}[1.cm]{\parbox{1.7cm}{\centerline{$d\sigma/dQ^2$} \centerline{(pb/GeV$^2$)}}}
                           & $\Delta_{stat}$ &
                             $\Delta_{syst}$ &
                             $\Delta_{ES}$ &
           \raisebox{0.2cm}[0.8cm]{\parbox{1.5cm}{\centerline{QED} \vspace{-.2cm} \centerline{correction}}
} &
{\raisebox{0.15cm}[1.cm]{{\parbox{1.5cm}{\centerline{$C_{\rm had}$}}}}} \\[.05cm]
\hline
$\;\,200\; - \;500$  & $0.0105$  & $\pm 0.0010$ & $_{-0.0010}^{+0.0012}$ & $_{-0.0011}^{+0.0014}$ & $1.045$ & $0.985$ \\
$\;\,500\; - \;1000$ & $0.0108$  & $\pm 0.0008$ & $_{-0.0005}^{+0.0007}$ & $_{-0.0004}^{+0.0004}$ & $1.033$ & $1.000$ \\
$1000\; - \;2000$    & $0.00571$  & $\pm 0.00042$ & $_{-0.00037}^{+0.00014}$ & $_{-0.00003}^{+0.00006}$ & $1.042$ & $1.001$ \\
$2000\; - \;4000$    & $0.00221$  & $\pm 0.00018$ & $_{-0.00008}^{+0.00005}$ & $_{-0.00004}^{+0.00007}$ & $1.064$ & $0.999$ \\
$4000\; - \;10000$   & $0.000380$ & $\pm 4.1\cdot 10^{-5}$ & $_{-1.1}^{+1.0}\cdot 10^{-5}$ & $_{-3.0}^{+3.6}\cdot 10^{-5}$ & $1.085$ & $0.998$ \\
\hline
\end{tabular}
\end{center}
\end{table} 

\begin{table}
\begin{center}
\begin{tabular}{|l|cccc||c|c|}
\hline
\multicolumn{7}{|c|}{1999-2000 $e^+p$ data sample ($\sqrt{s}=318$ GeV)}\\
\hline
\multicolumn{1}{|c|}{\raisebox{0.25cm}[1.cm]{\parbox{1.7cm}{\centerline{$Q^2$ range} \centerline{(GeV$^2$)}}}}
          & \raisebox{0.25cm}[1.cm]{\parbox{1.7cm}{\centerline{$d\sigma/dQ^2$} \centerline{(pb/GeV$^2$)}}}
                           & $\Delta_{stat}$ &
                             $\Delta_{syst}$ &
                             $\Delta_{ES}$ &
           \raisebox{0.2cm}[0.8cm]{\parbox{1.5cm}{\centerline{QED} \vspace{-.2cm} \centerline{correction}}
} & {\raisebox{0.15cm}[1.cm]{{\parbox{1.5cm}{\centerline{$C_{\rm had}$}}}}} \\[.05cm]
\hline
$\;\,200\; - \;500$  & $0.0125$ & $\pm 0.0010$ & $_{-0.0014}^{+0.0012}$ & $_{-0.0013}^{+0.0016}$ & $1.055$ & $0.983$ \\
$\;\,500\; - \;1000$ & $0.0107$ & $\pm 0.0007$ & $_{-0.0006}^{+0.0007}$ & $_{-0.0005}^{+0.0005}$ & $1.048$ & $0.999$ \\
$1000\; - \;2000$    & $0.00668$ & $\pm 0.00038$ & $_{-0.00014}^{+0.00010}$ & $_{-0.00007}^{+0.00009}$ & $1.054$ & $1.000$ \\
$2000\; - \;4000$    & $0.00233$ & $\pm 0.00016$ & $_{-0.00008}^{+0.00002}$ & $_{-0.00004}^{+0.00006}$ & $1.073$ & $0.998$ \\
$4000\; - \;10000$   & $0.000489$ & $\pm 4.0\cdot 10^{-5}$ & $_{-1.3}^{+0.9}\cdot 10^{-5}$ & $_{-3.3}^{+4.3}\cdot 10^{-5}$ & $1.097$ & $0.997$ \\
\hline
\end{tabular}
\end{center}
\end{table}  

\begin{table}
\begin{center}
\begin{tabular}{|l|cccc|c|}
\hline
\multicolumn{6}{|c|}{Combined 1995-2000 $e^+p$ data sample ($\sqrt{s}= 318$ GeV)}\\
\hline
\multicolumn{1}{|c|}{\raisebox{0.25cm}[1.cm]{\parbox{1.7cm}{\centerline{$Q^2$ range} \centerline{(GeV$^2$)}}}}
          & \raisebox{0.25cm}[1.cm]{\parbox{1.7cm}{\centerline{$d\sigma/dQ^2$} \centerline{(pb/GeV$^2$)}}}
                           & $\Delta_{stat}$ &
                             $\Delta_{syst}$ &
                             $\Delta_{ES}$ & \raisebox{0.25cm}[1.cm]{${\sigma_{318}^{th}}/{\sigma_{300}^{th}}$}\\
\hline
$\;\,200\; - \;500$ &$0.0119$  &$\pm 0.0007$ & $^{+0.0012}_{-0.0012}$ &$^{+0.0016}_{-0.0013}$ & 1.0476\\
$\;\,500\; - \;1000$&$0.0110$  &$\pm 0.0005$ & $^{+0.0007}_{-0.0005}$ &$^{+0.0005}_{-0.0004}$ & 1.0580\\
$1000\; - \;2000$   &$0.00647$ &$\pm 0.00029$ & $^{+0.00008}_{-0.00018}$&$^{+0.00008}_{-0.00005}$ & 1.0750\\
$2000\; - \;4000$   &$0.00237$ &$\pm 0.00012$ & $^{+0.00002}_{-0.00007}$&$^{+0.00007}_{-0.00005}$ & 1.1022\\
$4000\; - \;10000$  &$0.000472$&$\pm 3.1\cdot 10^{-5}$&$^{+0.7}_{-1.2}\cdot 10^{-5}$ & $^{+4.2}_{-3.4}\cdot 10^{-5}$ & 1.1633\\
\hline
\end{tabular}
\caption{
Inclusive jet cross-section $d\sigma/dQ^2$ for jets of hadrons in the laboratory frame. The statistical, systematic and energy-scale uncertainties are shown separately. The multiplicative correction applied to correct for QED radiative effects and for hadronisation effects and the theoretical correction factor used to combine the two data sets are shown.
}
  \label{table-q2incjet}
\end{center}
\end{table}  

\newpage

\begin{table}
\begin{center}
\begin{tabular}{|l|cccc||c|c|}
\hline
\multicolumn{7}{|c|}{1995-1997 $e^+p$ data sample ($\sqrt{s}=300$ GeV)}\\
\hline
\multicolumn{1}{|c|}{\raisebox{0.15cm}[1.cm]{\parbox{1.7cm}{\centerline{$\eta^{\rm jet}$ range}}}} &
\raisebox{0.25cm}[1.cm]{\parbox{1.7cm}{\centerline{$d\sigma/d\eta^{\rm jet}$} \centerline{(pb)}}} & $\Delta_{stat}$ & $\Delta_{syst}$ & $\Delta_{ES}$ &
           \raisebox{0.2cm}[0.8cm]{\parbox{1.5cm}{\centerline{QED} \vspace{-.2cm} \centerline{correction}}} & {\raisebox{0.15cm}[1.cm]{{\parbox{1.5cm}{\centerline{$C_{\rm had}$}}}}} \\[.05cm]
\hline 

$-1\;\; {\rm to} \;0$        & $3.80$ & $\pm 0.37$ & $_{-0.31}^{+0.25}$ & $_{-0.10}^{+0.14}$ & $1.082$ &$0.967$ \\
$\;\;0\;\;\,{\rm to}\;1$   & $9.60$ & $\pm 0.53$ &   $_{-0.51}^{+0.37}$ & $_{-0.05}^{+0.05}$ & $1.052$ &$0.992$ \\
$\;\;1\;\;\,{\rm to}\;1.5$ & $10.53$& $\pm 0.79$ &   $_{-0.38}^{+0.22}$ & $_{-0.04}^{+0.05}$ & $1.045$ &  $1.005$ \\
$\;1.5\; {\rm to} \;2$       & $7.78$ & $\pm 0.67$ & $_{-0.40}^{+0.55}$ & $_{-0.02}^{+0.04}$ & $1.042$ & $1.017$ \\
\hline
\end{tabular}
\end{center}
\end{table}

\begin{table}
\begin{center}
\begin{tabular}{|l|cccc||c|c|}
\hline
\multicolumn{7}{|c|}{1999-2000 $e^+p$ data sample ($\sqrt{s}=318$ GeV)$\;\;$}\\
\hline
\multicolumn{1}{|c|}{\raisebox{0.15cm}[1.cm]{\parbox{1.7cm}{\centerline{$\eta^{\rm jet}$ range}}}} &
\raisebox{0.25cm}[1.cm]{\parbox{1.7cm}{\centerline{$d\sigma/d\eta^{\rm jet}$} \centerline{(pb)}}} & $\Delta_{stat}$ & $\Delta_{syst}$ & $\Delta_{ES}$ &
           \raisebox{0.2cm}[0.8cm]{\parbox{1.5cm}{\centerline{QED} \vspace{-.2cm} \centerline{correction}}
} &
{\raisebox{0.15cm}[1.cm]{{\parbox{1.5cm}{\centerline{$C_{\rm had}$}}}}} \\[.05cm]
\hline 
$-1\;\; {\rm to} \;0$        & $4.76$ & $\pm 0.35$ & $_{-0.44}^{+0.11}$ & $_{-0.15}^{+0.20}$ & $1.086$ & $0.967$ \\
$\;\;0\;\;\,{\rm to}\;1$   & $9.73$ & $\pm 0.45$ &   $_{-0.23}^{+0.38}$ & $_{-0.04}^{+0.06}$ & $1.064$ & $0.991$ \\
$\;\;1\;\;\,{\rm to}\;1.5$ &$11.46$ & $\pm 0.69$ &   $_{-0.16}^{+0.19}$ & $_{-0.04}^{+0.06}$ & $1.058$ & $1.003$ \\
$\;1.5\; {\rm to}\;2$       & $9.55$ & $\pm 0.62$ &  $_{-0.29}^{+0.25}$ & $_{-0.04}^{+0.04}$ & $1.060$ & $1.013$ \\
\hline
\end{tabular}
\end{center}
\end{table}

\begin{table}
\begin{center}
\begin{tabular}{|l|cccc|c|}
\hline
\multicolumn{6}{|c|}{Combined 1995-2000 $e^+p$ data sample ($\sqrt{s}= 318$ GeV)}\\
\hline
\multicolumn{1}{|c|}{\raisebox{0.15cm}[1.cm]{\parbox{1.7cm}{\centerline{$\eta^{\rm jet}$ range}}}} &
\raisebox{0.25cm}[1.cm]{\parbox{1.7cm}{\centerline{$d\sigma/d\eta^{\rm jet}$} \centerline{(pb)}}} & $\Delta_{stat}$ & $\Delta_{syst}$ & $\Delta_{ES}$ & \raisebox{0.25cm}[1.cm]{${\sigma_{318}^{th}}/{\sigma_{300}^{th}}$}\\
\hline
$-1\;\; {\rm to} \;0$     & $4.48$&$\pm 0.26$&$^{+0.16}_{-0.39}$&$^{+0.18}_{-0.13}$ & $1.0670$\\
$\;\;0\;\;\,{\rm to}\;1$  & $9.98$&$\pm 0.35$&$^{+0.38}_{-0.31}$&$^{+0.06}_{-0.05}$ & $1.0797$ \\
$\;\;1\;\;\,{\rm to}\;1.5$&$11.50$&$\pm 0.54$&$^{+0.17}_{-0.19}$&$^{+0.06}_{-0.04}$ & $1.0990$ \\
$\;1.5\; {\rm to} \;2$    & $9.26$&$\pm 0.48$&$^{+0.32}_{-0.31}$&$^{+0.04}_{-0.03}$ & $1.1299$ \\
\hline
\end{tabular}
\caption{
Inclusive jet cross-section $d\sigma/d\eta^{\rm jet}$ for jets of hadrons in the laboratory frame. For details, see the caption to Table~\ref{table-q2incjet}.
}
  \label{table-etaincjet}
\end{center}
\end{table}  

\clearpage


\begin{table}
\begin{center}
\begin{tabular}{|l|cccc||c|c|}
\hline
\multicolumn{7}{|c|}{1995-1997 $e^+p$ data sample ($\sqrt{s}=300$ GeV)}\\
\hline
\multicolumn{1}{|c|}{\raisebox{0.25cm}[1.cm]{\parbox{1.7cm}{\centerline{$E_T^{\rm jet}$ range} \centerline{(GeV)}}}}
          & \raisebox{0.25cm}[1.cm]{\parbox{1.7cm}{\centerline{$d\sigma/dE_T^{\rm jet}$} \centerline{(pb/GeV)}}}
                           & $\Delta_{stat}$ &
                             $\Delta_{syst}$ &
                             $\Delta_{ES}$ &
           \raisebox{0.2cm}[0.8cm]{\parbox{1.5cm}{\centerline{QED} \vspace{-.2cm} \centerline{correction}}
} &
{\raisebox{0.15cm}[1.cm]{{\parbox{1.5cm}{\centerline{$C_{\rm had}$}}}}} \\[.05cm]
\hline
$14\; - \;21$  & $0.950$ & $\pm 0.068$ &     $_{-0.069}^{+0.081}$ &    $_{-0.014}^{+0.014}$ & $1.025$ & $0.992$ \\
$21\; - \;29$  & $0.703$ & $\pm 0.052$ &     $_{-0.051}^{+0.040}$ &    $_{-0.004}^{+0.006}$ & $1.036$ & $0.998$ \\
$29\; - \;41$  & $0.486$ & $\pm 0.034$ &     $_{-0.022}^{+0.008}$ &    $_{-0.004}^{+0.006}$ & $1.066$ & $0.998$ \\
$41\; - \;55$  & $0.219$ & $\pm 0.021$ &     $_{-0.007}^{+0.008}$ &    $_{-0.004}^{+0.004}$ & $1.094$ & $0.998$ \\
$55\; - \;71$  & $0.0542$ & $\pm 0.0096$ &   $_{-0.0026}^{+0.0026}$ &  $_{-0.0020}^{+0.0020}$ & $1.131$ & $0.988$ \\
$71\; - \;87$  & $0.0210$ & $\pm 0.0058$ &   $_{-0.0008}^{+0.0007}$ &  $_{-0.0012}^{+0.0011}$ & $1.147$ & $0.986$ \\
$87\; - \;119$ & $0.00643$ & $\pm 0.00214$ & $_{-0.00064}^{+0.00121}$ &$_{-0.00047}^{+0.00070}$ & $1.219$ & $0.981$ \\
\hline
\end{tabular}
\end{center}
\end{table}

\begin{table}
\begin{center}
\begin{tabular}{|l|cccc||c|c|}
\hline
\multicolumn{7}{|c|}{1999-2000 $e^+p$ data sample ($\sqrt{s}=318$ GeV)$\;\;$}\\
\hline
\raisebox{0.25cm}[1.cm]{\parbox{1.7cm}{\centerline{$E_T^{\rm jet}$ range} \centerline{(GeV)}}}
          & \raisebox{0.25cm}[1.cm]{\parbox{1.7cm}{\centerline{$d\sigma/dE_T^{\rm jet}$} \centerline{(pb/GeV)}}}
                           & $\Delta_{stat}$ &
                             $\Delta_{syst}$ &
                             $\Delta_{ES}$ &
           \raisebox{0.2cm}[0.8cm]{\parbox{1.5cm}{\centerline{QED} \vspace{-.2cm} \centerline{correction}}
} &
{\raisebox{0.15cm}[1.cm]{{\parbox{1.5cm}{\centerline{$C_{\rm had}$}}}}} \\[.05cm]
\hline
$14\; - \;21$ & $1.030$ & $\pm 0.060$  & $_{-0.092}^{+0.056}$ & $_{-0.012}^{+0.014}$ & $1.035$ & $0.989$\\
$21\; - \;29$ & $0.816$ & $\pm 0.047$  & $_{-0.034}^{+0.022}$ & $_{-0.007}^{+0.008}$ & $1.052$ & $0.995$\\
$29\; - \;41$ & $0.527$ & $\pm 0.030$  & $_{-0.007}^{+0.014}$ & $_{-0.005}^{+0.006}$ & $1.071$ & $0.998$\\
$41\; - \;55$ & $0.230$ & $\pm 0.018$  & $_{-0.006}^{+0.003}$ & $_{-0.004}^{+0.004}$ & $1.112$ & $0.998$\\
$55\; - \;71$ & $0.0775$& $\pm 0.0096$ & $_{-0.0035}^{+0.0026}$ & $_{-0.0026}^{+0.0026}$ & $1.135$ & $0.990$\\
$71\; - \;87$ &$0.0232$ & $\pm 0.0052$ & $_{-0.0012}^{+0.0010}$ & $_{-0.0012}^{+0.0012}$ & $1.172$ & $0.986$\\
$87\; - \;119$&$0.00385$& $\pm 0.00146$& $_{-0.00012}^{+0.00034}$ & $_{-0.00033}^{+0.00036}$ & $1.192$ & $0.984$\\
\hline
\end{tabular}
\end{center}
\end{table}

\begin{table}
\begin{center}
\begin{tabular}{|l|cccc|c|}
\hline
\multicolumn{6}{|c|}{Combined 1995-2000 $e^+p$ data sample ($\sqrt{s}= 318$ GeV)}\\
\hline
\multicolumn{1}{|c|}{\raisebox{0.25cm}[1.cm]{\parbox{1.7cm}{\centerline{$E_T^{\rm jet}$ range} \centerline{(GeV)}}}}
          & \raisebox{0.25cm}[1.cm]{\parbox{1.7cm}{\centerline{$d\sigma/dE_T^{\rm jet}$} \centerline{(pb/GeV)}}}
                           & $\Delta_{stat}$ &
                             $\Delta_{syst}$ &
                             $\Delta_{ES}$ & \raisebox{0.25cm}[1.cm]{${\sigma_{318}^{th}}/{\sigma_{300}^{th}}$}\\
\hline
$14\; - \;21$ &$1.022$  &$\pm 0.046$  &$^{+0.067}_{-0.084}$ &$^{+0.014}_{-0.013}$ & $1.0621$ \\
$21\; - \;29$ &$0.791$  &$\pm 0.036$  &$^{+0.027}_{-0.034}$ &$^{+0.007}_{-0.006}$ & $1.0688$\\
$29\; - \;41$ &$0.527$  &$\pm 0.023$  &$^{+0.009}_{-0.009}$ &$^{+0.006}_{-0.005}$ & $1.0855$\\
$41\; - \;55$ &$0.236$  &$\pm 0.014$  &$^{+0.004}_{-0.004}$ &$^{+0.004}_{-0.004}$ & $1.1216$\\
$55\; - \;71$ &$0.0727$ &$\pm 0.0074$ &$^{+0.0020}_{-0.0029}$&$^{+0.0025}_{-0.0025}$ & $1.1891$\\
$71\; - \;87$ &$0.0246$ &$\pm 0.0043$ &$^{+0.0010}_{-0.0008}$&$^{+0.0013}_{-0.0014}$ & $1.3046$\\
$87\; - \;119$&$0.00572$&$\pm 0.00143$&$^{+0.00066}_{-0.00027}$&$^{+0.00058}_{-0.00045}$ & $1.5461$\\
\hline
\end{tabular}
\caption{
Inclusive jet cross-section $d\sigma/dE_T^{\rm jet}$ for jets of hadrons in the laboratory frame. For details, see the caption to Table~\ref{table-q2incjet}.
}
  \label{table-etincjet}
\end{center}
\end{table}  

\newpage


\begin{table}
\begin{center}
\begin{tabular}{|l|cccc||c|c|}
\hline
\multicolumn{7}{|c|}{1995-1997 $e^+p$ data sample ($\sqrt{s}=300$ GeV)}\\
\hline
\multicolumn{1}{|c|}{\raisebox{0.25cm}[1.cm]{\parbox{1.7cm}{\centerline{$Q^2$ range} \centerline{(GeV$^2$)}}}}
          & \raisebox{0.25cm}[1.cm]{\parbox{1.7cm}{\centerline{$d\sigma/dQ^2$} \centerline{(pb/GeV$^2$)}}}
                           & $\Delta_{stat}$ &
                             $\Delta_{syst}$ &
                             $\Delta_{ES}$ &
           \raisebox{0.2cm}[0.8cm]{\parbox{1.5cm}{\centerline{QED} \vspace{-.2cm} \centerline{correction}}
} &
{\raisebox{0.15cm}[1.cm]{{\parbox{1.5cm}{\centerline{$C_{\rm had}$}}}}} \\[.05cm]
\hline
$\;\,200\; - \;500$ & $0.00153$ & $\pm 0.00051$ & $_{-0.00067}^{+0.00030}$ & 
$_{-0.00023}^{+0.00031}$ & $1.040$ & $0.916$ \\
$\;\,500\; - \;1000$& $0.00165$ & $\pm 0.00034$ & $_{-0.00022}^{+0.00015}$ & 
$_{-0.00013}^{+0.00014}$ & $1.045$ & $0.924$ \\
$1000\; - \;2000$   & $0.00112$ & $\pm 0.00020$ & $_{-0.00007}^{+0.00006}$ & 
$_{-0.00003}^{+0.00003}$ & $1.059$ & $0.926$ \\
$2000\; - \;4000$   & $3.85\cdot 10^{-4}$ & $\pm 0.82 \cdot 10^{-4}$ & $_{-0.26}^{+0.25}\cdot 10^{-4}$ & $_{-0.15}^{+0.26}\cdot 10^{-4}$ & $1.077$ & $0.910$ \\
$4000\; - \;10000$  & $7.13\cdot 10^{-5}$ & $\pm 1.91 \cdot 10^{-5}$ & $_{-0.12}^{+0.70}\cdot 10^{-5}$ & $_{-0.79}^{+0.77}\cdot 10^{-5}$ & $1.083$ & $0.893$ \\
\hline
\end{tabular}
\end{center}
\end{table}

\begin{table}
\begin{center}
\begin{tabular}{|l|cccc||c|c|}
\hline
\multicolumn{7}{|c|}{1999-2000 $e^+p$ data sample ($\sqrt{s}=318$ GeV)}\\
\hline
\multicolumn{1}{|c|}{\raisebox{0.25cm}[1.cm]{\parbox{1.7cm}{\centerline{$Q^2$ range} \centerline{(GeV$^2$)}}}}
          & \raisebox{0.25cm}[1.cm]{\parbox{1.7cm}{\centerline{$d\sigma/dQ^2$} \centerline{(pb/GeV$^2$)}}}
                           & $\Delta_{stat}$ &
                             $\Delta_{syst}$ &
                             $\Delta_{ES}$ &
           \raisebox{0.2cm}[0.8cm]{\parbox{1.5cm}{\centerline{QED} \vspace{-.2cm} \centerline{correction}}
} &
{\raisebox{0.15cm}[1.cm]{{\parbox{1.5cm}{\centerline{$C_{\rm had}$}}}}} \\[.05cm]
\hline
$\;\,200\; - \;500$ & $0.00290$ & $\pm 0.00062$ & $_{-0.00079}^{+0.00081}$ & 
$_{-0.00043}^{+0.00057}$ & $1.068$ & $0.914$ \\
$\;\,500\; - \;1000$& $0.00190$ & $\pm 0.00031$ & $_{-0.00026}^{+0.00023}$ & 
$_{-0.00015}^{+0.00017}$ & $1.054$ & $0.925$ \\
$1000\; - \;2000$   & $0.00112$ & $\pm 0.00016$ & $_{-0.00007}^{+0.00006}$ & 
$_{-0.00003}^{+0.00004}$ & $1.067$ & $0.927$ \\
$2000\; - \;4000$   & $4.02\cdot 10^{-4}$ & $\pm 0.70\cdot 10^{-4}$ & $_{-0.32}^{+0.27}\cdot 10^{-4}$ & 
$_{-0.15}^{+0.16}\cdot 10^{-4}$ & $1.078$ & $0.916$ \\
$4000\; - \;10000$  & $11.70\cdot 10^{-5}$ & $\pm 2.04\cdot 10^{-5}$ & $_{-0.99}^{+1.12}\cdot 10^{-5}$ & 
$_{-0.99}^{+1.32}\cdot 10^{-5}$ & $1.101$ & $0.904$ \\
\hline
\end{tabular}
\end{center}
\end{table}

\begin{table}
\begin{center}
\begin{tabular}{|l|cccc|c|}
\hline
\multicolumn{6}{|c|}{Combined 1995-2000 $e^+p$ data sample ($\sqrt{s}= 318$ GeV)}\\
\hline
\multicolumn{1}{|c|}{\raisebox{0.25cm}[1.cm]{\parbox{1.7cm}{\centerline{$Q^2$ range} \centerline{(GeV$^2$)}}}}
          & \raisebox{0.25cm}[1.cm]{\parbox{1.7cm}{\centerline{$d\sigma/dQ^2$} \centerline{(pb/GeV$^2$)}}}
                           & $\Delta_{stat}$ &
                             $\Delta_{syst}$ &
                             $\Delta_{ES}$ & \raisebox{0.25cm}[1.cm]{${\sigma_{318}^{th}}/{\sigma_{300}^{th}}$}\\
\hline
$\;\,200\; - \;500$ &$0.00242$&$\pm 0.00043$&$^{+0.00050}_{-0.00069}$ & $^{+0.00048}_{-0.00036}$ &$1.0848$\\
$\;\,500\; - \;1000$&$0.00186$&$\pm 0.00024$&$^{+0.00019}_{-0.00024}$ &$^{+0.00016}_{-0.00015}$ &$1.0859$\\
$1000\; - \;2000$   &$0.00116$&$\pm 0.00013$&$^{+0.00006}_{-0.00005}$ &$^{+0.00004}_{-0.00003}$ &$1.0926$\\
$2000\; - \;4000$   &$4.13\cdot 10^{-4}$&$\pm 0.56 \cdot 10^{-4}$ &$^{+0.25}_{-0.28}\cdot 10^{-4}$ &$^{+0.21}_{-0.16}\cdot 10^{-4}$ &$1.1191$\\
$4000\; - \;10000$  &$10.49\cdot 10^{-5}$&$\pm  1.53 \cdot 10^{-5}$ &$^{+0.98}_{-0.57}\cdot 10^{-5}$ &$^{+1.17}_{-0.97}\cdot 10^{-5}$ &$1.1777$\\
\hline
\end{tabular}
\caption{
Dijet cross-section $d\sigma/dQ^2$ for jets of hadrons in the laboratory frame. For details, see the caption to Table~\ref{table-q2incjet}.
}
  \label{table-q2dijet}
\end{center}
\end{table}  


\begin{table}
\begin{center}
\begin{tabular}{|c|cccc||c|c|}
\hline
\multicolumn{7}{|c|}{1995-1997 $e^+p$ data sample ($\sqrt{s}=300$ GeV)}\\
\hline
\multicolumn{1}{|c|}{\raisebox{0.25cm}[1.cm]{\parbox{1.7cm}{\centerline{$m_{12}$ range} \centerline{(GeV)}}}}
          & \raisebox{0.25cm}[1.cm]{\parbox{1.7cm}{\centerline{$d\sigma/dm_{12}$} \centerline{(pb/GeV)}}}
                           & $\Delta_{stat}$ &
                             $\Delta_{syst}$ &
                             $\Delta_{ES}$ &
           \raisebox{0.2cm}[0.8cm]{\parbox{1.5cm}{\centerline{QED} \vspace{-.2cm} \centerline{correction}}
} &
{\raisebox{0.15cm}[1.cm]{{\parbox{1.5cm}{\centerline{$C_{\rm had}$}}}}} \\[.05cm]
\hline
$10\; - \;20$ & $0.0724$& $\pm 0.0137$& $_{-0.0084}^{+0.0030}$ & $_{-0.0011}^{+0.0019}$ & $1.050$ & $0.921$ \\
$20\; - \;30$ & $0.104$ & $\pm 0.018$ & $_{-0.007}^{+0.006}$ & $_{-0.003}^{+0.002}$ & $1.053$ & $0.901$ \\
$30\; - \;40$ & $0.0884$& $\pm 0.0184$& $_{-0.0092}^{+0.0098}$ & $_{-0.0029}^{+0.0038}$ & $1.060$ & $0.918$ \\
$40\; - \;75$ & $0.0296$& $\pm 0.0068$& $_{-0.0046}^{+0.0054}$ & $_{-0.0014}^{+0.0015}$ & $1.089$ & $0.942$ \\
\hline
\end{tabular}
\end{center}
\end{table}  

\begin{table}
\begin{center}
\begin{tabular}{|c|cccc||c|c|}
\hline
\multicolumn{7}{|c|}{1999-2000 $e^+p$ data sample ($\sqrt{s}=318$ GeV)}\\
\hline
\multicolumn{1}{|c|}{\raisebox{0.25cm}[1.cm]{\parbox{1.7cm}{\centerline{$m_{12}$ range} \centerline{(GeV)}}}} & \raisebox{0.25cm}[1.cm]{\parbox{1.7cm}{\centerline{$d\sigma/dm_{12}$} \centerline{(pb/GeV)}}}
                           & $\Delta_{stat}$ &
                             $\Delta_{syst}$ &
                             $\Delta_{ES}$ &
           \raisebox{0.2cm}[0.8cm]{\parbox{1.5cm}{\centerline{QED} \vspace{-.2cm} \centerline{correction}}
} &
{\raisebox{0.15cm}[1.cm]{{\parbox{1.5cm}{\centerline{$C_{\rm had}$}}}}} \\[.05cm]
\hline
$10\; - \;20$ & $0.0913$ & $\pm 0.0129$ & $_{-0.0045}^{+0.0094}$ & $_{-0.0019}^{+0.0016}$ & $1.059$ & $0.918$ \\
$20\; - \;30$ & $0.140$ & $\pm 0.018$   & $_{-0.011}^{+0.008}$   & $_{-0.003}^{+0.004}$ & $1.064$ & $0.902$ \\
$30\; - \;40$ & $0.110$ & $\pm 0.018$ & $_{-0.009}^{+0.010}$ & $_{-0.004}^{+0.005}$ & $1.082$ & $0.917$ \\
$40\; - \;75$ & $0.0228$ & $\pm 0.0049$ & $_{-0.0021}^{+0.0045}$ & $_{-0.0011}^{+0.0010}$ & $1.085$ & $0.941$ \\
\hline
\end{tabular}
\end{center}
\end{table}  

\begin{table}
\begin{center}
\begin{tabular}{|l|cccc|c|}
\hline
\multicolumn{6}{|c|}{Combined 1995-2000 $e^+p$ data sample ($\sqrt{s}= 318$ GeV)}\\
\hline
\multicolumn{1}{|c|}{\raisebox{0.25cm}[1.cm]{\parbox{1.7cm}{\centerline{$m_{12}$ range} \centerline{(GeV)}}}}
          & \raisebox{0.25cm}[1.cm]{\parbox{1.7cm}{\centerline{$d\sigma/dm_{12}$} \centerline{(pb/GeV)}}} & $\Delta_{stat}$ & $\Delta_{syst}$ & $\Delta_{ES}$ & \raisebox{0.25cm}[1.cm]{${\sigma_{318}^{th}}/{\sigma_{300}^{th}}$}\\
\hline
$10\; - \;20$&$0.0863$&$\pm 0.0098$&$^{+0.0058}_{-0.0057}$&$^{+0.0018}_{-0.0016}$&$1.0830$ \\
$20\; - \;30$&$0.130$& $\pm 0.013$ &$^{+0.006}_{-0.009}$&  $^{+0.003}_{-0.003}$&$1.0992$ \\
$30\; - \;40$&$0.106$&$\pm 0.013$&$^{+0.009}_{-0.008}$&$^{+0.005}_{-0.003}$&$1.1172$ \\
$40\; - \;75$&$0.0270$&$\pm 0.0042$&$^{+0.0051}_{-0.0031}$&$^{+0.0013}_{-0.0013}$&$1.1552$ \\
\hline
\end{tabular}
\caption{
Dijet cross-section $d\sigma/dm_{12}$ for jets of hadrons in the laboratory frame. For details, see the caption to Table~\ref{table-q2incjet}.
}
\label{table-m12dijet}
\end{center}
\end{table}  


\begin{table}[p]
\begin{center}
\begin{tabular}{|c|ccc||c|}
\hline
\raisebox{0.25cm}[1.cm]{\parbox{2cm}{\centerline{$y_{cut}$ value}}}
          & \raisebox{0.25cm}[1.cm]{\parbox{2cm}{\centerline{$\big< n_{sbj} \big>$}}}
                           & \raisebox{0.25cm}[1.cm]{$\Delta_{stat}$} &
                             \raisebox{0.25cm}[1.cm]{$\Delta_{syst}$} &
  {\raisebox{0.15cm}[1.cm]{{\parbox{1.5cm}{\centerline{$C_{\rm had}$}}}}} \\[.05cm]
\hline
\hline
\multicolumn{5}{|c|}{$14<E_T^{\rm jet}<17$ GeV} \\
\hline
   $0.0005$ & $4.432$ & $\pm 0.082$ &  $^{+0.078}_{-0.025}$ &  $2.127$
   \\[.2cm]
   $0.001$  & $3.576$ & $\pm 0.071$ &  $^{+0.098}_{-0.020}$ &  $1.845$
   \\[.2cm]
   $0.003$  & $2.616$ & $\pm 0.057$ &  $^{+0.058}_{-0.020}$ &  $1.526$
   \\[.2cm]
   $0.005$  & $2.238$ & $\pm 0.050$ &  $^{+0.039}_{-0.022}$ &  $1.426$
   \\[.2cm]
   $0.01$   & $1.755$ & $\pm 0.042$ &  $^{+0.026}_{-0.031}$ &  $1.292$
   \\[.2cm]
   $0.03$   & $1.263$ & $\pm 0.033$ &  $^{+0.028}_{-0.015}$ &  $1.094$
   \\[.2cm]
   $0.05$   & $1.117$ & $\pm 0.023$ &  $^{+0.019}_{-0.013}$ &  $1.037$
   \\[.2cm]
   $0.1$    & $1.010$ & $\pm 0.007$ &  $^{+0.018}_{-0.003}$ &  $1.005$
   \\[.2cm]

\hline
\hline
\multicolumn{5}{|c|}{$17<E_T^{\rm jet}<21$  GeV} \\
\hline
   $0.0005$ & $4.272$ & $\pm 0.071$ &  $^{+0.067}_{-0.050}$ &  $1.989$
   \\[.2cm]
   $0.001$  & $3.522$ & $\pm 0.060$ &  $^{+0.058}_{-0.043}$ &  $1.726$
   \\[.2cm]
   $0.003$  & $2.427$ & $\pm 0.047$ &  $^{+0.049}_{-0.025}$ &  $1.464$
   \\[.2cm]
   $0.005$  & $1.999$ & $\pm 0.042$ &  $^{+0.036}_{-0.012}$ &  $1.364$
   \\[.2cm]
   $0.01$   & $1.574$ & $\pm 0.034$ &  $^{+0.021}_{-0.017}$ &  $1.227$
   \\[.2cm]
   $0.03$   & $1.176$ & $\pm 0.026$ &  $^{+0.014}_{-0.018}$ &  $1.054$
   \\[.2cm]
   $0.05$   & $1.102$ & $\pm 0.020$ &  $^{+0.021}_{-0.010}$ &  $1.016$
   \\[.2cm]
   $0.1$    & $1.021$ & $\pm 0.009$ &  $^{+0.006}_{-0.006}$ &  $0.999$
   \\[.2cm]
\hline
\end{tabular}
\caption{
Mean subjet multiplicity as a function of $y_{\rm cut}$ for the $E_T^{\rm jet}$ regions $14<E_T^{\rm jet}<17$ and $17<E_T^{\rm jet}<21$  GeV.
The statistical
and systematic uncertainties are shown separately. The multiplicative correction
applied to correct for hadronisation effects is shown
in the last column.
}
  \label{table-sbjycut1}
\end{center}
\end{table}

\begin{table}[p]
\begin{center}
\begin{tabular}{|c|ccc||c|}
\hline
\raisebox{0.25cm}[1.cm]{\parbox{2cm}{\centerline{$y_{cut}$ value}}}
          & \raisebox{0.25cm}[1.cm]{\parbox{2cm}{\centerline{$\big< n_{sbj} \big>$}}}
                           & \raisebox{0.25cm}[1.cm]{$\Delta_{stat}$} &
                             \raisebox{0.25cm}[1.cm]{$\Delta_{syst}$} &
  {\raisebox{0.15cm}[1.cm]{{\parbox{1.5cm}{\centerline{$C_{\rm had}$}}}}} \\[.05cm]
\hline
\hline
\multicolumn{5}{|c|}{$21<E_T^{\rm jet}<25$ GeV} \\
\hline
   $0.0005$ & $4.051$ & $\pm 0.070$ &  $^{+0.067}_{-0.010}$ &  $1.857$
   \\[.2cm]
   $0.001$  & $3.251$ & $\pm 0.059$ &  $^{+0.071}_{-0.011}$ &  $1.637$
   \\[.2cm]
   $0.003$  & $2.266$ & $\pm 0.043$ &  $^{+0.052}_{-0.014}$ &  $1.409$
   \\[.2cm]
   $0.005$  & $1.953$ & $\pm 0.038$ &  $^{+0.042}_{-0.008}$ &  $1.304$
   \\[.2cm]
   $0.01$   & $1.504$ & $\pm 0.034$ &  $^{+0.035}_{-0.007}$ &  $1.167$
   \\[.2cm]
   $0.03$   & $1.164$ & $\pm 0.025$ &  $^{+0.023}_{-0.008}$ &  $1.025$
   \\[.2cm]
   $0.05$   & $1.070$ & $\pm 0.017$ &  $^{+0.023}_{-0.005}$ &  $1.003$
   \\[.2cm]
   $0.1$    & $1.022$ & $\pm 0.009$ &  $^{+0.008}_{-0.005}$ &  $0.997$
   \\[.2cm]
\hline
\hline
\multicolumn{5}{|c|}{$25<E_T^{\rm jet}<35$ GeV} \\
\hline
   $0.0005$ & $3.786$ & $\pm 0.049$ &  $^{+0.027}_{-0.021}$ &  $1.736$
   \\[.2cm]
   $0.001$  & $2.997$ & $\pm 0.039$ &  $^{+0.028}_{-0.011}$ &  $1.570$
   \\[.2cm]
   $0.003$  & $2.067$ & $\pm 0.029$ &  $^{+0.017}_{-0.009}$ &  $1.339$
   \\[.2cm]
   $0.005$  & $1.717$ & $\pm 0.025$ &  $^{+0.019}_{-0.012}$ &  $1.228$
   \\[.2cm]
   $0.01$   & $1.386$ & $\pm 0.023$ &  $^{+0.018}_{-0.009}$ &  $1.099$
   \\[.2cm]
   $0.03$   & $1.137$ & $\pm 0.016$ &  $^{+0.005}_{-0.012}$ &  $1.006$
   \\[.2cm]
   $0.05$   & $1.061$ & $\pm 0.011$ &  $^{+0.003}_{-0.013}$ &  $0.998$
   \\[.2cm]
   $0.1$    & $1.008$ & $\pm 0.004$ &  $^{+0.004}_{-0.004}$ &  $0.998$
   \\[.2cm]
\hline
\end{tabular}
\caption{
Mean subjet multiplicity as a function of $y_{\rm cut}$ for the $E_T^{\rm jet}$ regions $21<E_T^{\rm jet}<25$ and $25<E_T^{\rm jet}<35$ GeV.  For details, see the caption to Table~\ref{table-sbjycut1}.
}
  \label{table-sbjycut2}
\end{center}
\end{table}

\begin{table}[p]
\begin{center}
\begin{tabular}{|c|ccc||c|}
\hline
\raisebox{0.25cm}[1.cm]{\parbox{2cm}{\centerline{$y_{cut}$ value}}}
          & \raisebox{0.25cm}[1.cm]{\parbox{2cm}{\centerline{$\big< n_{sbj} \big>$}}}
                           & \raisebox{0.25cm}[1.cm]{$\Delta_{stat}$} &
                             \raisebox{0.25cm}[1.cm]{$\Delta_{syst}$} &
  {\raisebox{0.15cm}[1.cm]{{\parbox{1.5cm}{\centerline{$C_{\rm had}$}}}}} \\[.05cm]
\hline
\hline
\multicolumn{5}{|c|}{$35<E_T^{\rm jet}<55$ GeV} \\
\hline
   $0.0005$ & $3.343$ & $\pm 0.045$ &  $^{+0.037}_{-0.017}$ &  $1.579$
   \\[.2cm]
   $0.001$  & $2.649$ & $\pm 0.040$ &  $^{+0.038}_{-0.013}$ &  $1.433$
   \\[.2cm]
   $0.003$  & $1.797$ & $\pm 0.029$ &  $^{+0.029}_{-0.009}$ &  $1.197$
   \\[.2cm]
   $0.005$  & $1.512$ & $\pm 0.026$ &  $^{+0.027}_{-0.006}$ &  $1.106$
   \\[.2cm]
   $0.01$   & $1.245$ & $\pm 0.022$ &  $^{+0.020}_{-0.004}$ &  $1.026$
   \\[.2cm]
   $0.03$   & $1.106$ & $\pm 0.015$ &  $^{+0.003}_{-0.002}$ &  $0.996$
   \\[.2cm]
   $0.05$   & $1.060$ & $\pm 0.011$ &  $^{+0.005}_{-0.001}$ &  $0.996$
   \\[.2cm]
   $0.1$    & $1.010$ & $\pm 0.004$ &  $^{+0.001}_{-0.000}$ &  $0.998$
   \\[.2cm]
\hline
\hline
\multicolumn{5}{|c|}{$55<E_T^{\rm jet}<119$ GeV} \\
\hline
   $0.0005$ & $2.790$ & $\pm 0.068$ &  $^{+0.013}_{-0.035}$ &  $1.450$
   \\[.2cm]
   $0.001$  & $2.196$ & $\pm 0.056$ &  $^{+0.022}_{-0.026}$ &  $1.285$
   \\[.2cm]
   $0.003$  & $1.520$ & $\pm 0.041$ &  $^{+0.038}_{-0.012}$ &  $1.073$
   \\[.2cm]
   $0.005$  & $1.378$ & $\pm 0.042$ &  $^{+0.040}_{-0.015}$ &  $1.022$
   \\[.2cm]
   $0.01$   & $1.260$ & $\pm 0.039$ &  $^{+0.024}_{-0.012}$ &  $0.998$
   \\[.2cm]
   $0.03$   & $1.098$ & $\pm 0.025$ &  $^{+0.005}_{-0.007}$ &  $0.996$
   \\[.2cm]
   $0.05$   & $1.044$ & $\pm 0.017$ &  $^{+0.005}_{-0.003}$ &  $0.998$
   \\[.2cm]
   $0.1$    & $1.036$ & $\pm 0.026$ &  $^{+0.004}_{-0.002}$ &  $1.000$
   \\[.2cm]
\hline
\end{tabular}
\caption{
Mean subjet multiplicity as a function of $y_{\rm cut}$ for the $E_T^{\rm jet}$ regions $35<E_T^{\rm jet}<55$ and $55<E_T^{\rm jet}<119$ GeV.
For details, see the caption to Table~\ref{table-sbjycut1}.
}
  \label{table-sbjycut3}
\end{center}
\end{table}

\begin{table}[p]
\begin{center}
\begin{tabular}{|l|ccc||c|}
\hline
\raisebox{0.25cm}[1.cm]{\parbox{2.6cm}{\centerline{$Q^2$ range}\centerline{(GeV$^2$)}}}
          & \raisebox{0.25cm}[1.cm]{\parbox{2cm}{\centerline{$\big< n_{sbj} \big>$}}} 
                           & \raisebox{0.25cm}[1.cm]{$\Delta_{stat}$} &
                             \raisebox{0.25cm}[1.cm]{$\Delta_{syst}$} &
  {\raisebox{0.15cm}[1.cm]{{\parbox{1.5cm}{\centerline{$C_{\rm had}$}}}}} \\[.05cm]
\hline
   $200\;\;\,-\;350$    & $1.695$ & $\pm 0.053$ & $^{+0.025}_{-0.027}$ & $1.316$ \\[.2cm]
   $350\;\;\,-\;500$    & $1.677$ & $\pm 0.050$ & $^{+0.011}_{-0.049}$ & $1.257$ \\[.2cm]
   $500\;\;\,-\;750$    & $1.516$ & $\pm 0.036$ & $^{+0.028}_{-0.011}$ & $1.212$ \\[.2cm]
   $750\;\;\,-\;1000$   & $1.487$ & $\pm 0.040$ & $^{+0.042}_{-0.018}$ & $1.175$ \\[.2cm]
   $1000\;-\;2000$      & $1.391$ & $\pm 0.024$ & $^{+0.030}_{-0.012}$ & $1.129$ \\[.2cm]
   $2000\;-\;4000$      & $1.372$ & $\pm 0.027$ & $^{+0.012}_{-0.014}$ & $1.081$ \\[.2cm]
   $4000\;-\;10000$     & $1.318$ & $\pm 0.031$ & $^{+0.034}_{-0.010}$ & $1.051$ \\[.2cm]

 \hline
\end{tabular}
\caption{
Measurement of the mean subjet multiplicity at $y_{\rm cut}=10^{-2}$
as a function of $Q^2$.
For details, see the caption to Table~\ref{table-sbjycut1}.
}
  \label{table-sbq2}
\end{center}
\end{table}
%


\begin{figure}
\vspace*{-.2cm}
  \unitlength 1cm   
\vfill
  \begin{picture}(17,17)
    \put(0.,0.){\epsfig{file=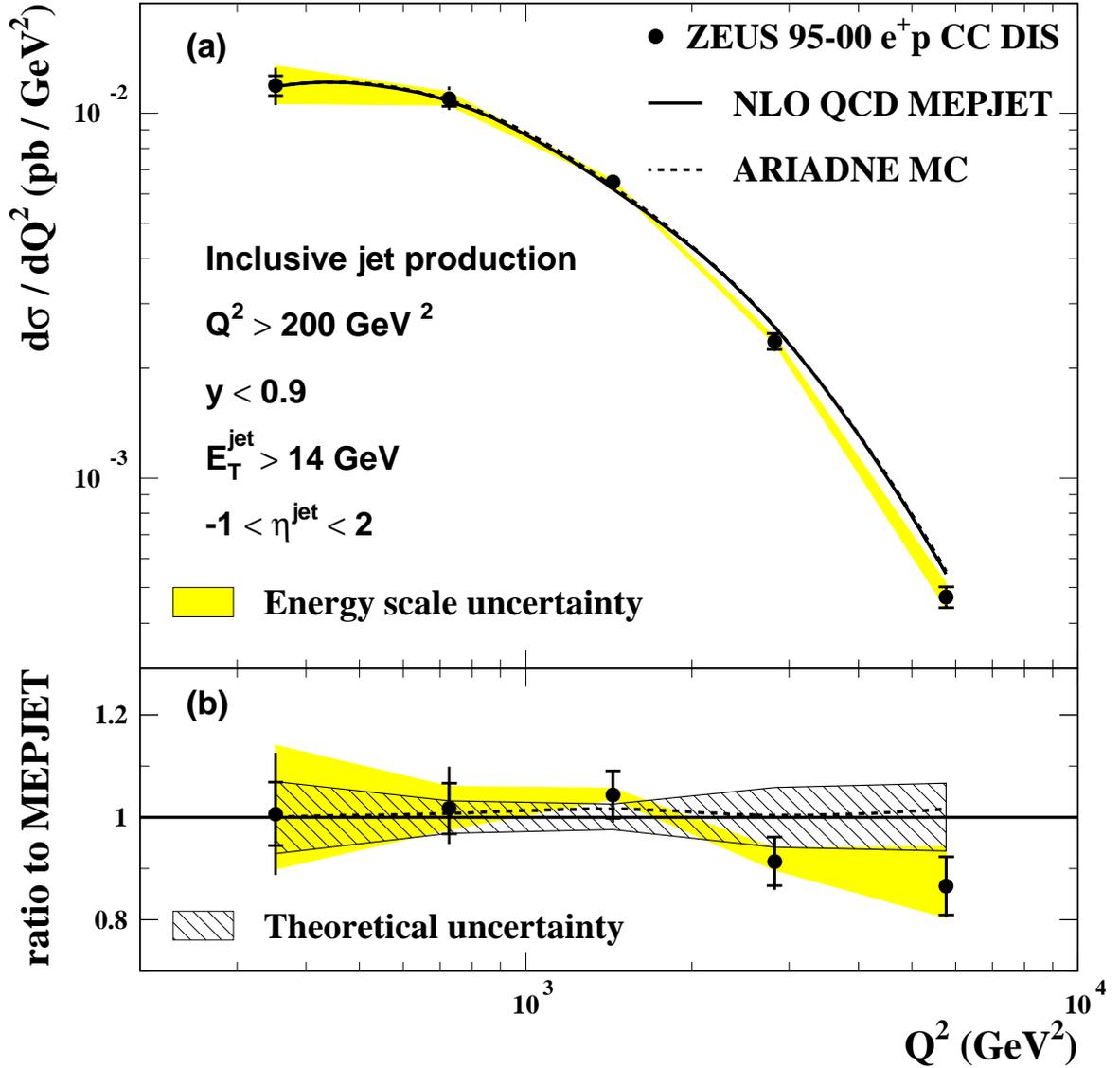,width=17.cm,angle=0,clip=}}
  \end{picture}
\vspace*{-1.cm}
\caption{(a) The differential cross-section $d\sigma/dQ^2$ for inclusive jet production in the laboratory frame with $E_T^{jet} > 14$ GeV and $-1<\eta^{jet}<2$  in the kinematic region $Q^2>200$ GeV~$^2$ and $y<0.9$ for the 1995-2000 $e^+p$ data (black dots). The data are corrected to hadron level. The inner error bars represent the statistical uncertainty of the data, the outer error bars show the statistical and the systematic uncertainties (not associated with the uncertainty in the absolute energy scale) added in quadrature. The shaded band displays the uncertainty due to the absolute energy scale of the CAL. 
The parton shower Monte Carlo prediction given by ARIADNE at hadron level (dashed line) and the next-to-leading-order prediction obtained with MEPJET corrected to hadron level (solid line) are shown. 
(b)~The ratio of the measured cross section to the next-to-leading-order calculation. The theoretical uncertainty is indicated by the hatched band.
}
\label{fig-incq2}
\vfill
\end{figure}

\begin{figure}
  \unitlength 1cm   
\vfill
  \begin{picture}(17,17)
    \put(0.,0.){\epsfig{file=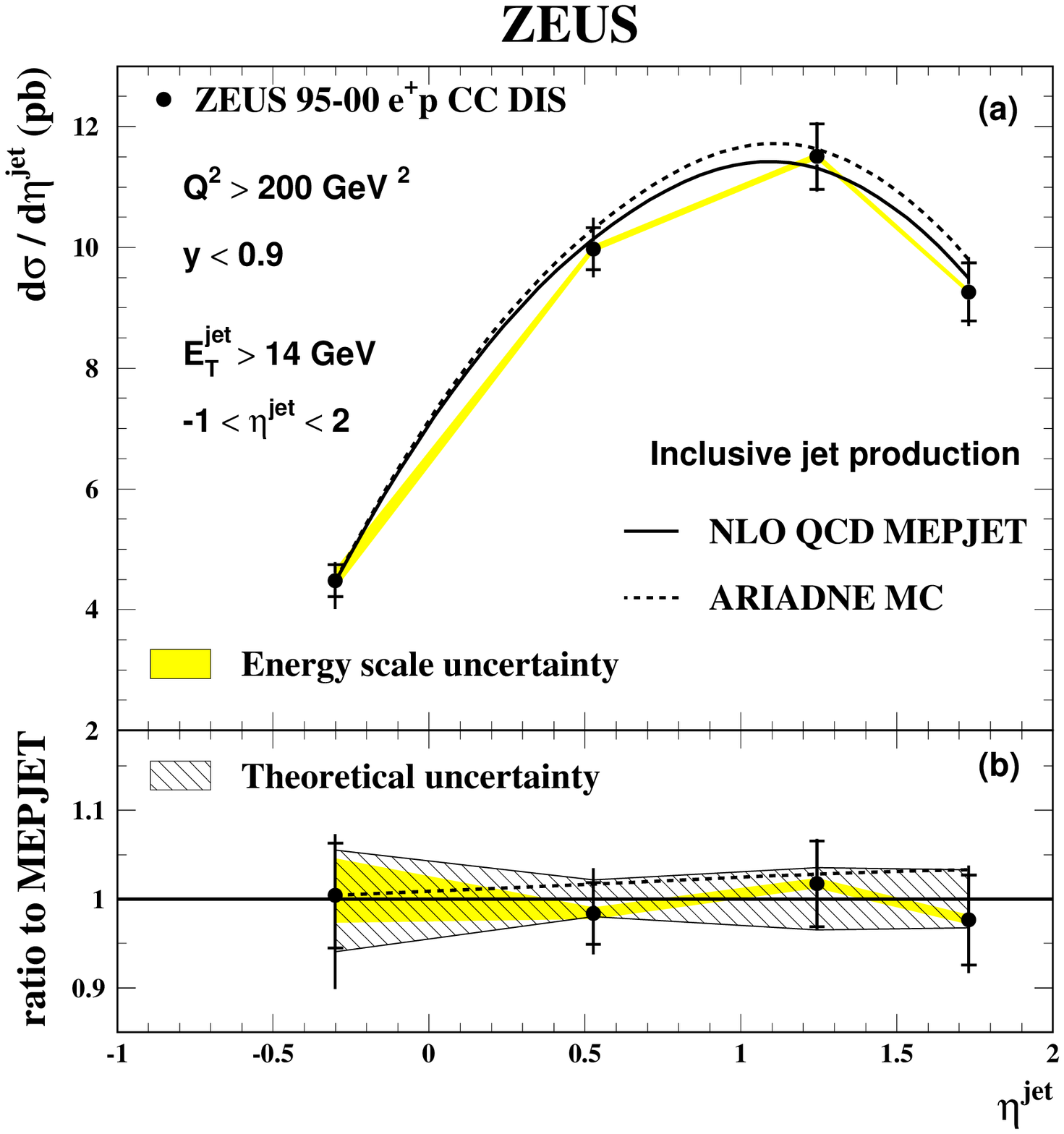,width=17.cm,angle=0,clip=}}
  \end{picture}
\caption{(a) The differential cross-section $d\sigma/d\eta^{jet}$ for inclusive jet production in the laboratory frame with $E_T^{jet} > 14$ GeV and $-1<\eta^{jet}<2$  in the kinematic region $Q^2>200$ GeV $^2$ and $y<0.9$. 
Other details are as decribed in the caption to Fig.~\ref{fig-incq2}.
}
\label{fig-inceta}
\vfill
\end{figure}

\begin{figure}
  \unitlength 1cm   
\vfill
  \begin{picture}(17,17)
    \put(0.,0.){\epsfig{file=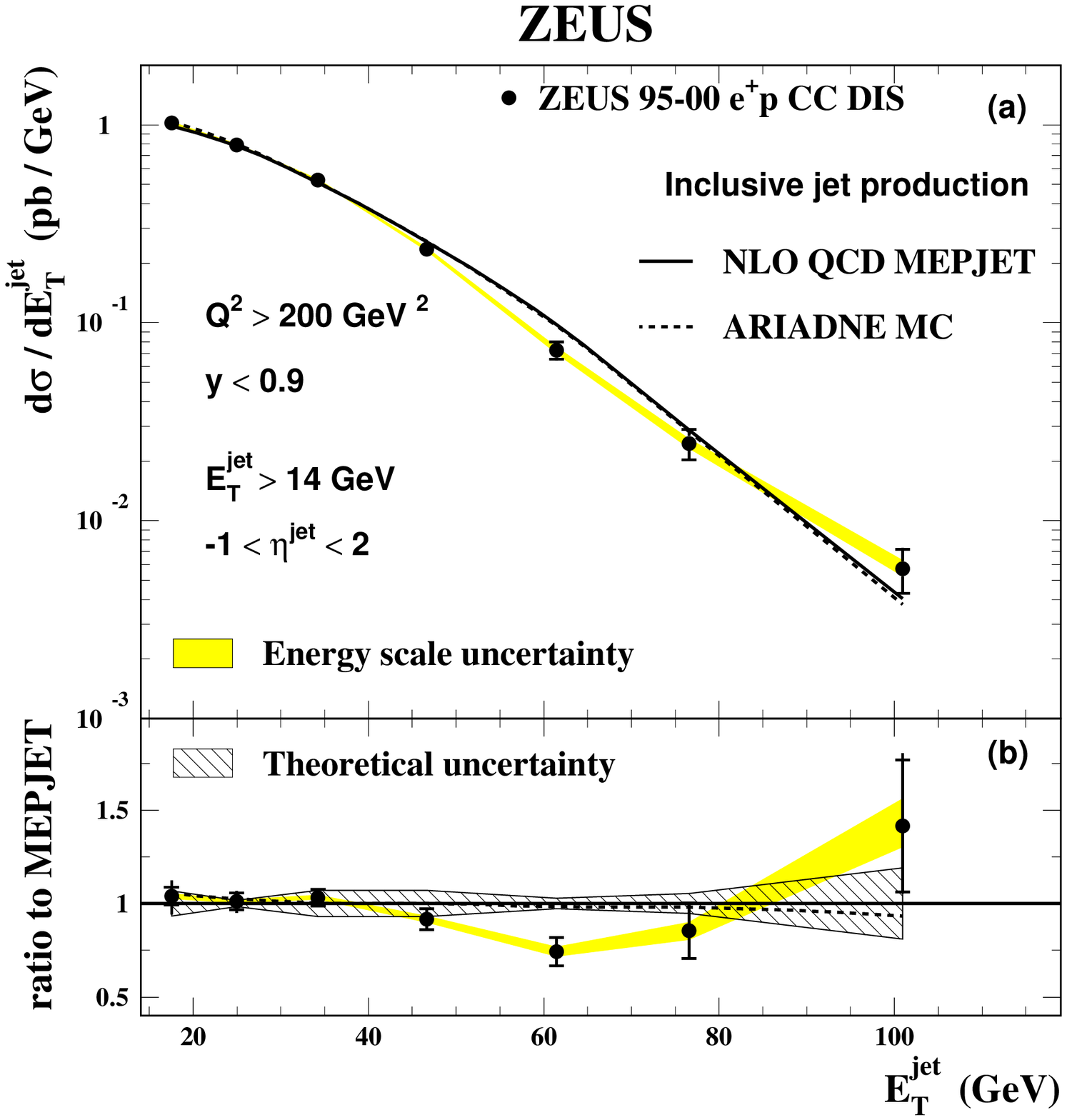,width=17.cm,angle=0,clip=}}
  \end{picture}
\caption{(a) The differential cross-section $d\sigma/dE_T^{jet}$ for inclusive jet production in the laboratory frame with $E_T^{jet} > 14$ GeV and $-1<\eta^{jet}<2$  in the kinematic region $Q^2>200$ GeV $^2$ and $y<0.9$.  Other details are as decribed in the caption to Fig.~\ref{fig-incq2}.
}
\label{fig-incet}
\vfill
\end{figure}

\begin{figure}
  \unitlength 1cm   
\vfill
  \begin{picture}(17,17)
    \put(0.,0.){\epsfig{file=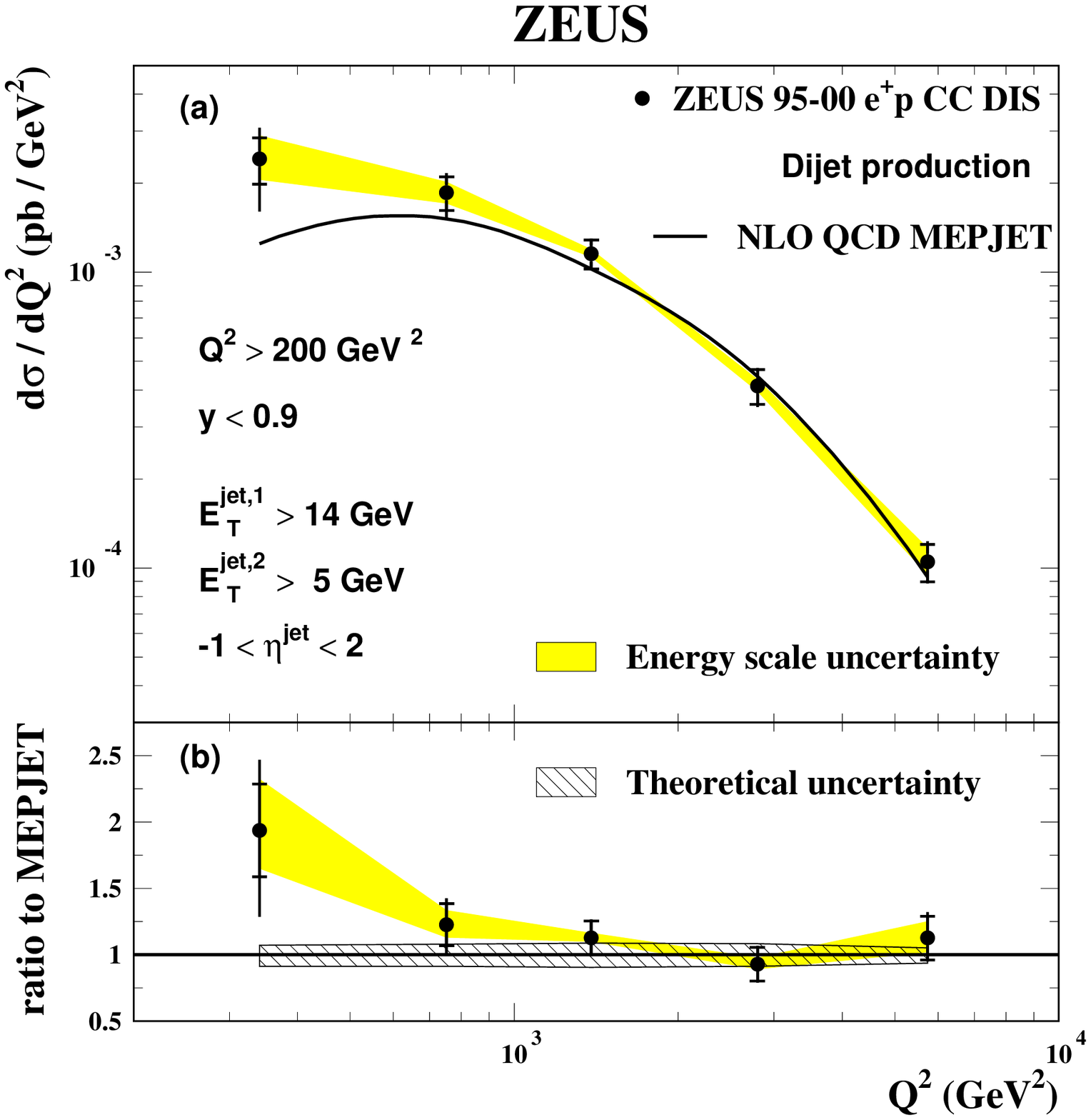,width=17.cm,angle=0,clip=}}
  \end{picture}
\caption{(a) The differential cross-section $d\sigma/dQ^2$ for dijet production in the laboratory frame with $E_T^{jet,1} > 14$ GeV, $E_T^{jet,2} > 5$ GeV and $-1<\eta^{jet}<2$  in the kinematic region $Q^2>200$ GeV $^2$ and $y<0.9$. Other details are as decribed in the caption to Fig.~\ref{fig-incq2}.
}
\label{fig-diq2}
\vfill
\end{figure}

\vfill

\begin{figure}
  \unitlength 1cm   
\vfill
  \begin{picture}(17,17)
    \put(0.,0.){\epsfig{file=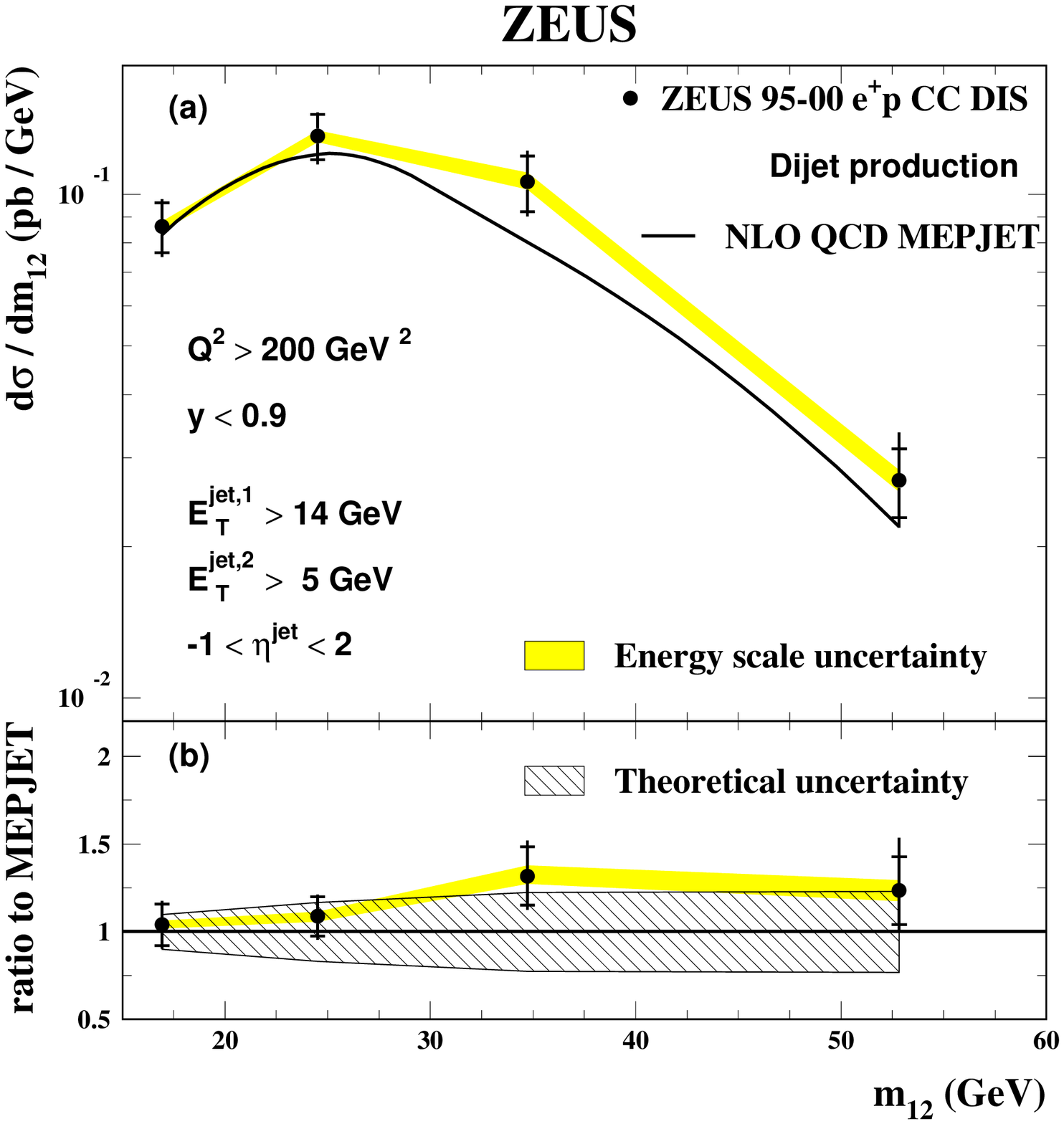,width=17.cm,angle=0,clip=}}
  \end{picture}
\caption{(a) The differential cross-section $d\sigma/dm_{12}$ for dijet production in the laboratory frame with $E_T^{jet,1} > 14$ GeV, $E_T^{jet,2} > 5$ GeV and $-1<\eta^{jet}<2$  in the kinematic region $Q^2>200$ GeV $^2$ and $y<0.9$. Other details are as decribed in the caption to Fig.~\ref{fig-incq2}.
}
\label{fig-dim12}
\vfill
\end{figure}

\begin{figure}
  \unitlength 1cm   
\vfill
  \begin{picture}(17,17)
    \put(0.,0.){\epsfig{file=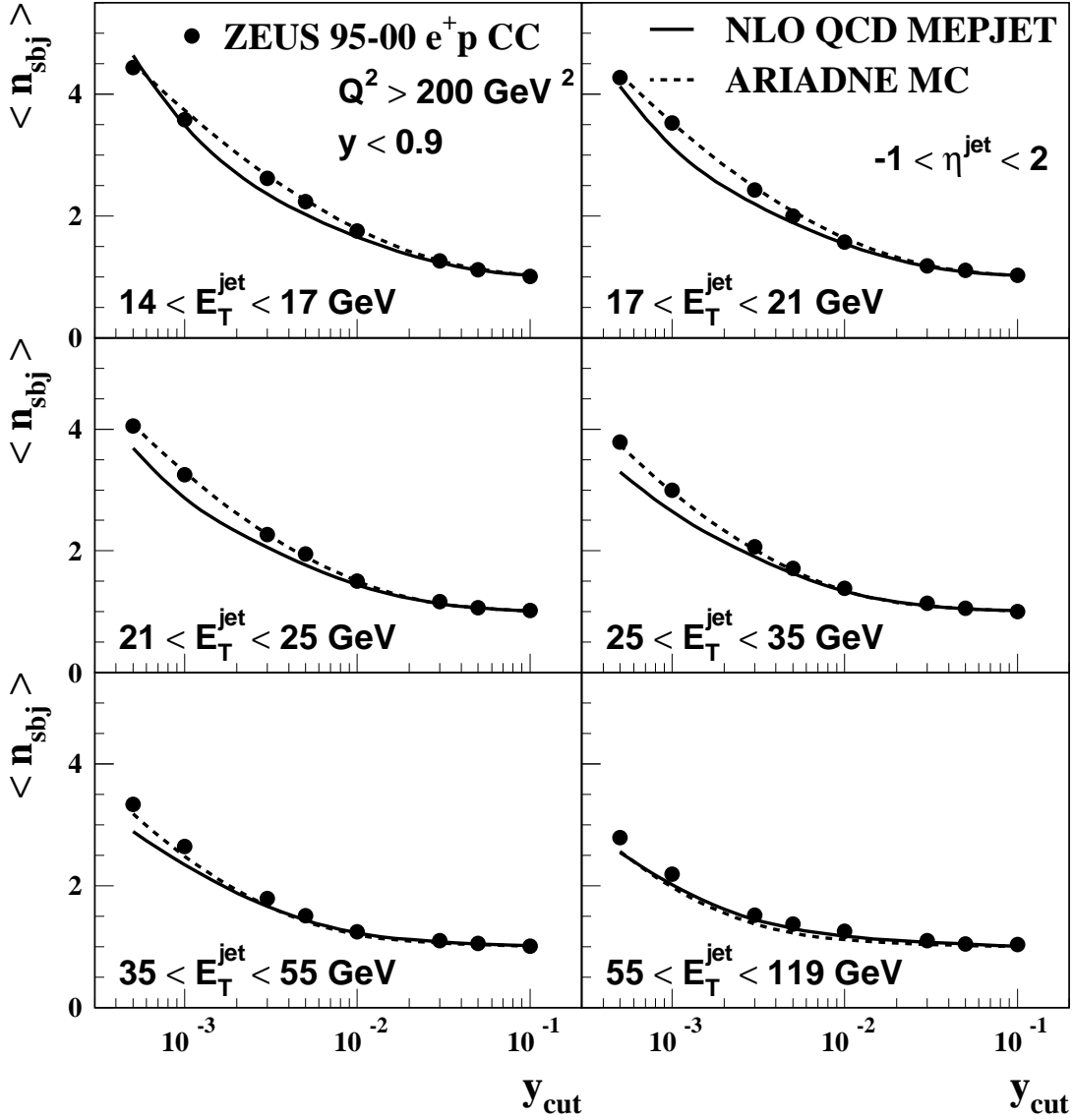,width=17.cm,angle=0,clip=}}
  \end{picture}
\caption{Mean subjet multiplicity (black dots), $\langle n_{sbj}\rangle$, as a function of $y_{cut}$ for inclusive jet production in the laboratory frame with $-1< \eta^{jet} <2$ in different $E_T^{jet}$ regions. 
The parton shower Monte Carlo prediction given by ARIADNE at hadron level (dashed line) and the next-to-leading-order prediction obtained with MEPJET corrected to hadron level (solid line) are shown. The error bars are smaller than the symbols.
}
\label{fig-subyet}
\vfill
\end{figure}

\begin{figure}
  \unitlength 1cm   
\vfill
  \begin{picture}(17,17)
    \put(0.,0.){\epsfig{file=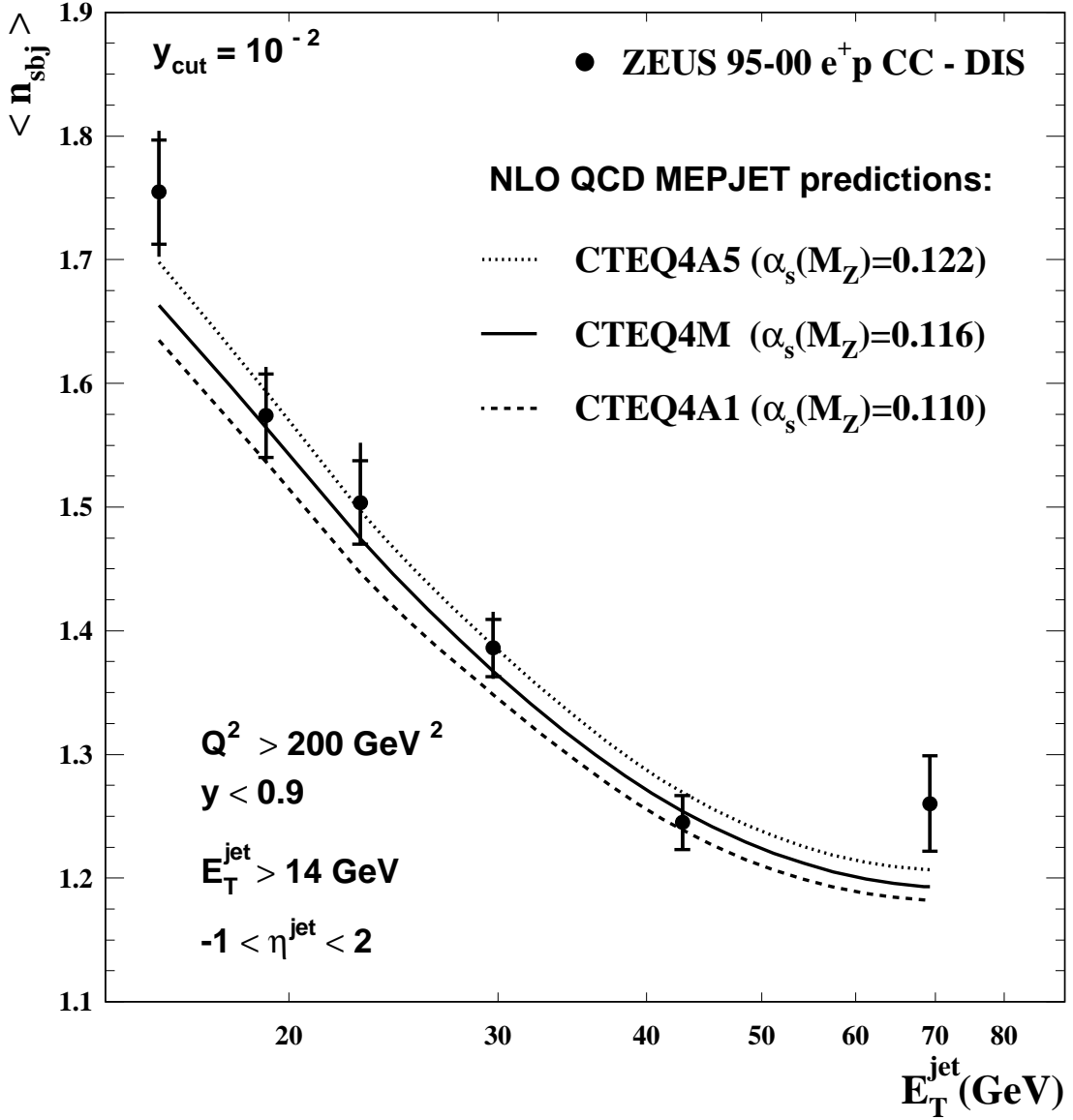,width=17.cm,angle=0,clip=}}
  \end{picture}
\caption{Mean subjet multiplicity, $\langle n_{sbj}\rangle$, at $y_{cut}=10^{-2}$ as a function of $E_T^{jet}$ (black dots), for inclusive jet production  in the laboratory frame with $E_T^{jet} > 14$ GeV and $-1< \eta^{jet} <2$. The inner error bars represent the statistical uncertainty of the data. The outer error bars show the statistical and systematic uncertainties added in quadrature. The NLO QCD predictions obtained with MEPJET using the CTEQ4 sets of proton PDFs are shown for 3 different values of $\als(M_Z)$ (curves). 
}
\label{fig-alphassent}
\vfill
\end{figure}

\begin{figure}
  \unitlength 1cm   
\vfill
  \begin{picture}(16.,16.)
    \put(0.,0.){\epsfig{file=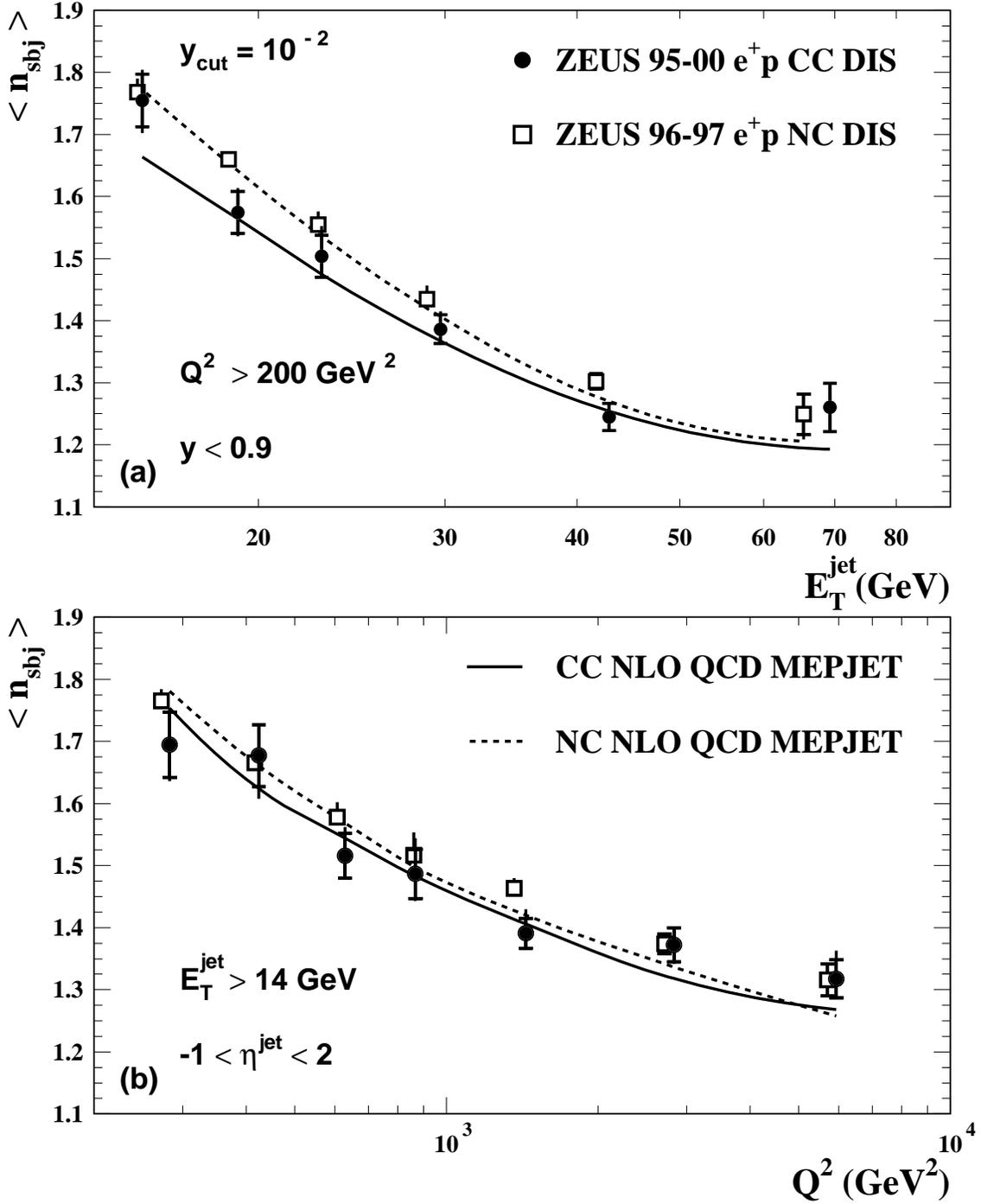,width=17.cm,angle=0,clip=}}
 \end{picture}
\caption{
Measurements of  \nsub at $y_{cut}=10^{-2}$ for inclusive jet production in the laboratory frame with $E_T^{jet} > 14$ GeV and $-1< \eta^{jet} <2$ in charged current DIS (circles) and neutral current DIS (open squares) as a function of (a) $E_T^{jet}$ and (b) $Q^2$.
}
\label{fig-subyetnc}
\vfill
\end{figure}

\end{document}